\documentclass[epj]{svjour}
\usepackage{graphics}
\usepackage{amsmath,amssymb}
\newcommand{\tanb}{\ensuremath{\tan\beta}}
\newcommand{\mZero}{\ensuremath{\mathrm{m}_0}}
\newcommand{\mOneHalf}{\ensuremath{\mathrm{m}_{1/2}}}
\newcommand{\AZero}{\ensuremath{\mathrm{A}_{0}}}
\newcommand{\MHOne}{\ensuremath{\mathrm{M}_{\mathrm{H}_1}}}
\newcommand{\MHTwo}{\ensuremath{\mathrm{M}_{\mathrm{H}_2}}}
\newcommand{\Mi}{\ensuremath{\mathrm{M}_i}}
\newcommand{\MOne}{\ensuremath{\mathrm{M}_1}}
\newcommand{\MTwo}{\ensuremath{\mathrm{M}_2}}
\newcommand{\MThree}{\ensuremath{\mathrm{M}_3}}
\newcommand{\MselL}{\ensuremath{\mathrm{M}_{\tilde{\mathrm{e}}_{\mathrm{L}}}}}
\newcommand{\MselR}{\ensuremath{\mathrm{M}_{\tilde{\mathrm{e}}_{\mathrm{R}}}}}
\newcommand{\MsmuL}{\ensuremath{\mathrm{M}_{\tilde{\mu}_{\mathrm{L}}}}}
\newcommand{\MsmuR}{\ensuremath{\mathrm{M}_{\tilde{\mu}_{\mathrm{R}}}}}
\newcommand{\MstauL}{\ensuremath{\mathrm{M}_{\tilde{\tau}_{\mathrm{L}}}}}
\newcommand{\MstauR}{\ensuremath{\mathrm{M}_{\tilde{\tau}_{\mathrm{R}}}}}
\newcommand{\MsqL}{\ensuremath{\mathrm{M}_{\tilde{\mathrm{q}}_{\mathrm{L}}}}}
\newcommand{\MsqR}{\ensuremath{\mathrm{M}_{\tilde{\mathrm{q}}_{\mathrm{R}}}}}
\newcommand{\MsquR}{\ensuremath{\mathrm{M}_{\tilde{\mathrm{u}}_{\mathrm{R}}}}}
\newcommand{\MsqdR}{\ensuremath{\mathrm{M}_{\tilde{\mathrm{d}}_{\mathrm{R}}}}}
\newcommand{\MsqsR}{\ensuremath{\mathrm{M}_{\tilde{\mathrm{s}}_{\mathrm{R}}}}}
\newcommand{\MsqcR}{\ensuremath{\mathrm{M}_{\tilde{\mathrm{c}}_{\mathrm{R}}}}}
\newcommand{\MsqOneL}{\ensuremath{\mathrm{M}_{{\tilde{\mathrm{q}}1}_{\mathrm{L}}}}}
\newcommand{\MsqTwoL}{\ensuremath{\mathrm{M}_{{\tilde{\mathrm{q}}2}_{\mathrm{L}}}}}
\newcommand{\MsqThreeL}{\ensuremath{\mathrm{M}_{{\tilde{\mathrm{q}}3}_{\mathrm{L}}}}}
\newcommand{\MstR}{\ensuremath{\mathrm{M}_{\tilde{\mathrm{t}}_{\mathrm{R}}}}}
\newcommand{\MsbR}{\ensuremath{\mathrm{M}_{\tilde{\mathrm{b}}_{\mathrm{R}}}}}
\newcommand{\Atop}{\ensuremath{\mathrm{A}_{t}}}
\newcommand{\Atau}{\ensuremath{\mathrm{A}_{\tau}}}
\newcommand{\Abottom}{\ensuremath{\mathrm{A}_{b}}}

\newcommand{\theTrace}{\ensuremath{\mathrm{Tr} [Y m^2]}}
\newcommand{\be}{\begin{equation}}
\newcommand{\ee}{\end{equation}}
\newcommand{\gev}{\ensuremath\mathrm{GeV}}
\begin{document}
\title{Measuring Unification}
\author{Claire Adam\inst{1} 
\and Jean-Loic Kneur\inst{2}
\and R\'emi Lafaye\inst{3}
\and Tilman Plehn\inst{4}
\and Michael Rauch\inst{5}
\and Dirk Zerwas\inst{1}
}                     
%
%
\institute{LAL, IN2P3/CNRS, Orsay, France
\and LPTA, Universit\'e Montpellier II, IN2P3/CNRS, Montpellier, France
\and LAPP, Universit\'e Savoie, IN2P3/CNRS, Annecy, France
\and Institut f\"ur Theoretische Physik, Universit\"at Heidelberg, Germany
\and Institut f\"ur Theoretische Physik, 
        Karlsruhe Institute of Technology, Karlsruhe, Germany
}
\date{October 26, 2010}
%
\abstract{ If supersymmetry is observed at the LHC its model
  parameters can be measured at the electroweak scale.  We discuss the
  expected precision on the parameter determination, including a
  proper treatment of experimental and theoretical errors.  Particular
  attention is paid to degenerate solutions.  Using the SFitter
  framework we perform a bottom-up reconstruction of the unified
  parameters at the high scale, including a full error
  propagation.
%
} 
\maketitle
\section{Introduction}
\label{intro}

While supersymmetry was originally not introduced as a
phenomenological model targeted at definite shortcomings of our
Standard Model, it has since developed into the most attractive
ultraviolet completion of the Standard Model at and above the TeV
scale.  The basis of supersymmetric theories is a matching of
fermionic degree of freedom with bosonic degree of freedom, now given
the particle content of the Standard Model at the electroweak
scale~\cite{Nath:2010zj,review,Martin:1997ns}. This symmetry automatically cures
the theoretical problem with a perturbatively unstable fundamental
Higgs mass in the Standard Model, i.e. the hierarchy problem.

Several specific points make the minimal supersymmetric extension of
the Standard Model (MSSM) an attractive ultraviolet completion: while
electroweak precision data favors a light Higgs
boson~\cite{Alcaraz:2007ri}, supersymmetry provides such a lightest
Higgs boson with a mass less than about 140~GeV. To implement
fundamental symmetries protecting the proton life time and avoiding
flavor-changing neutral currents into the MSSM we usually resort to
$R$ parity. This parity in turn forces the lightest supersymmetric
particle to be stable. Attributed to a `WIMP miracle' the typical
relic densities predicted for a weakly interacting supersymmetric dark
matter candidate roughly agree with the observed
values~\cite{dm_review1,dm_review2}.

In addition to these attractive weak-scale properties supersymmetry
offers another, unique, opportunity: it allows us to predictively
extrapolate a perturbative and renormalizable gauge theory to
arbitrarily high energy scales.  While we know that the three gauge
couplings do not unify in the Standard Model, including supersymmetric
degrees of freedom at the TeV scale can naturally lead to such a
unification~\cite{angle1,angle2,angle3,angle4,angle5,Marciano:1981un,angle6}. This is
one of the reasons why fundamental and unbroken supersymmetry could
live above the unification scale $Q_\text{GUT} >
10^{16}$~GeV~\cite{msugra1,msugra2,msugra3,msugra4,msugra5,msugra6,msugra7,msugra8,msugra9}. One
way of breaking supersymmetry would then be a gravity-driven link
between a hidden sector and the MSSM, where the scales get adjusted to
accommodate supersymmetry-breaking mass parameters around the TeV
scale. Using this setup a renormalization group analysis shows that
the electroweak symmetry is automatically broken by running one of the
two Higgs masses squared to negative values slightly below the mass
scale of the supersymmetric partners~\cite{Nilles:1982ik}.

In the coming years supersymmetry should be discovered at the LHC.  An
e$^+$e$^-$ linear collider (ILC) with a center--of--mass energy of
500~GeV extendible to 1~TeV, as it is under study, can shed further
light on supersymmetry.  A formidable task will be to determine the
fundamental parameters of supersymmetry from experimental
measurements. Even if we expect supersymmetry to unify at
a high scale we should not assume such a unification and simply fit
the high-scale parameters to experimental measurements at the
electroweak scale. Instead, the appropriate though technically
challenging question should be: 
do the measured weak-scale MSSM parameters evolved to high energy scales unify?
Or, if not, can we observe relics of a unification, like sum rules
stable with respect to renormalization group running~\cite{sumrules}.

In principle, grand unification of the supersymmetric parameters can
be observed from data in the gaugino and scalar
sectors~\cite{Blair:2002pg,Allanach:2004ud,Kneur:2008ur} or in the
gaugino sector only~\cite{Altunkaynak:2009tg}. The aspect we focus on
in this study is the determination of the
central values and errors~\cite{Lafaye:2004cn,Lafaye:2007vs}, i.e. a proper measurement
of supersymmetric unification. The bottom-up renormalization group
analysis are performed for expected LHC measurements
and its combination with the ILC.  Because our analysis
depends critically on the knowledge of all errors, we use the well
studied parameter set SPS1a~\cite{Allanach:2002nj}. It is interesting
to note that the result of the fit of the electroweak data, adding
b--physics observables, the anomalous magnetic moment of the
muon~\cite{Davier:2009zi} and the relic density~\cite{Larson:2010gs},
yields a best--fit point not too far from SPS1a, a further motivation
to study in detail such a parameter
choice~\cite{Buchmueller:2007zk,Bechtle:2009ty,Buchmueller:2008qe}.
Note that by adjusting slightly the parameters of SPS1a (denoted 
SPS1a' or SPA1 in~\cite{AguilarSaavedra:2005pw}) the relic density can be reduced 
to agree with the measurement from WMAP~\cite{Larson:2010gs} without changing the 
collider observables significantly. The relic density is only a side
product of our analysis, therefore we stick to SPS1a for which the
experimental error estimates are available directly. 

Our analysis relies on the SFitter
framework~\cite{Lafaye:2004cn,Lafaye:2007vs}.  The same techniques
have been successfully applied to other questions, like the
determination of the Higgs boson couplings at the
LHC~\cite{Lafaye:2009vr}.  The theoretical aspects of evolving the
relevant parameters from the electroweak to the GUT scale and
vice-versa we discussed in Section~\ref{sec:theo}.  The expected
measurements at the LHC and ILC are listed in Section~\ref{sec:data},
followed by a summary of the determination of the MSSM parameters at
the electroweak scale in Section~\ref{sec:paramEWscale}.  In
Section~\ref{sec:extrapol} we develop the method for the determination
of the unified parameters at the high scale and apply it to the
respective LHC and LHC+ILC measurements.

\section{Theoretical aspects}
\label{sec:theo}

For the following studies, the MSSM is defined as a variant of the phenomenological
MSSM with the following parameters
(evaluated for convenience at the electroweak scale, 1~TeV~\cite{AguilarSaavedra:2005pw}):
\tanb\ is the ratio of the vacuum expectation values of the Higgs doublets,
\MOne, \MTwo\ and \MThree\ are the gaugino mass parameters, $\mu$ is the higgsino
supersymmetric mass parameter and $m_A$ the physical (pole) mass of the pseudo scalar Higgs boson.  
For simplicity the gaugino masses are restricted to be positive.
The soft breaking masses in the slepton sector are denoted as
\MselL, \MselR, \MsmuL, \MsmuR, \MstauL\ and \MstauR. 
For the squarks, as u, d, s and c quarks are experimentally practically indistinguishable, a common definition (average) 
is used for the first two generations denoted as \MsqL\ and \MsqR. The third generation is treated separately
with the parameters \MsqThreeL, \MstR\ and \MsbR. 
The tri-linear terms, irrelevant for the first two generations due to the smallness of the fermion masses,
are taken into account for the third 
generation with \Atau, \Atop\ and \Abottom.

The supersymmetric and soft-supersymmetry breaking parameters of the MSSM 
can in principle be defined at some arbitrary scale, not only the electroweak scale. In order to
compare the results of the bottom-up and top-down approach, the high scale MSSM will also be used. 
The parameters of this model are defined at the GUT scale ($\sim 10^{16}$~GeV).  
In this model the two weak scale 
parameters $\mu$ and $m_A$ are replaced with the soft supersymmetry-breaking terms for the Higgs doublets, \MHOne\ and \MHTwo.

\subsection{Renormalization Group Evolution properties}
\label{sec:RGE}

\begin{table*}
\begin{center}
\begin{tabular}{llrrrrr}
\hline\noalign{\smallskip}
\multicolumn{2}{c}{ type of } & 
 \multicolumn{1}{c}{ nominal } & 
 \multicolumn{1}{c}{ stat. } & 
 \multicolumn{1}{c}{ LES } & 
 \multicolumn{1}{c}{ JES } & 
 \multicolumn{1}{c}{ theo. } \\
\multicolumn{2}{c}{ measurement } & 
 \multicolumn{1}{c}{ value } & 
 \multicolumn{4}{c}{ error } \\
\noalign{\smallskip}\hline\noalign{\smallskip}
$m_h$ & 
 & 108.7 & 0.01 & 0.25 &      & 2.0 \\
$m_t$ & 
 & 171.20& 0.01 &      & 1.0  &     \\
$m_{\tilde{l}_L}-m_{\chi_1^0}$ & 
 & 102.38& 2.3  & 0.1  &      & 1.1 \\
$m_{\tilde{g}}-m_{\chi_1^0}$ & 
 & 511.38& 2.3  &      & 6.0  & 6.1 \\
$m_{\tilde{q}_R}-m_{\chi_1^0}$ & 
 & 446.39& 10.0 &      & 4.3  & 5.5 \\
$m_{\tilde{g}}-m_{\tilde{b}_1}$ & 
 & 89.01 & 1.5  &      & 1.0  & 8.0 \\
$m_{\tilde{g}}-m_{\tilde{b}_2}$ & 
 & 62.93 & 2.5  &      & 0.7  & 8.2 \\
$m_{ll}^\mathrm{max}$: & three-particle edge($\chi_2^0$,$\tilde{l}_R$,$\chi_1^0$)  
 & 80.852 & 0.042& 0.08 &      & 1.2 \\
$m_{llq}^\mathrm{max}$: & three-particle edge($\tilde{q}_L$,$\chi_2^0$,$\chi_1^0$)  
 & 449.08& 1.4  &      & 4.3  & 5.1 \\
$m_{lq}^\mathrm{low}$: & three-particle edge($\tilde{q}_L$,$\chi_2^0$,$\tilde{l}_R$)
 & 326.32& 1.3  &      & 3.0  & 5.2 \\
$m_{ll}^\mathrm{max}(\chi_4^0)$: & three-particle edge($\chi_4^0$,$\tilde{l}_L$,$\chi_1^0$)
 & 277.36 & 3.3  & 0.3  &      & 2.0 \\
$m_{\tau\tau}^\mathrm{max}$: & three-particle edge($\chi_2^0$,$\tilde{\tau}_1$,$\chi_1^0$)
 & 83.21 & 5.0  &      & 0.8  & 1.0 \\
$m_{lq}^\mathrm{high}$: & four-particle edge($\tilde{q}_L$,$\chi_2^0$,$\tilde{l}_R$,$\chi_1^0$)
 & 390.18& 1.4  &      & 3.8  & 5.0 \\
$m_{llq}^\mathrm{thres}$: & threshold($\tilde{q}_L$,$\chi_2^0$,$\tilde{l}_R$,$\chi_1^0$)
 & 216.00& 2.3  &      & 2.0  & 3.3 \\
$m_{llb}^\mathrm{thres}$: & threshold($\tilde{b}_1$,$\chi_2^0$,$\tilde{l}_R$,$\chi_1^0$)
 & 198.41& 5.1  &      & 1.8  & 3.1 \\
\noalign{\smallskip}\hline
\end{tabular}
\end{center} \vspace*{-3mm}
\caption[]{
LHC measurements in SPS1a, taken 
  from~\cite{Weiglein:2004hn}. Shown are the nominal values (from SUSPECT~\cite{Djouadi:2002ze}) 
  and statistical errors, systematic errors from the lepton (LES)
  and jet energy scale (JES) and theoretical errors. 
  All values are given in GeV.}
\label{tab:LHCobs}
\end{table*}

The renormalization group equations (RGE)~\cite{RGE2} play an important role in the analysis 
by relating low to high scale MSSM parameters. 
Corrections up to the level of two--loops are implemented in SUSPECT~\cite{Djouadi:2002ze} and SoftSUSY~\cite{Allanach:2001kg}. 
Three-loop results for the beta functions are known and have been shown 
to stay within the two-loop error bands~\cite{3loop1,3loop2}. Unless stated otherwise
the full two-loop corrections are applied in the following. An approximate 
analytical form is useful to understand the RGE dependence of the most relevant parameters.
Typically, at leading one-loop order the RGE for the gaugino mass parameters are closely related
to the ones for the gauge coupling~\cite{RGE2}:
\begin{equation}
\frac{d}{dt}(\ln \Mi(t)) = \frac{B_i}{8\pi^2} g^2_i = \frac{d}{dt} (\ln g^2_i)\;,
\end{equation}
where $t=\ln Q$, $B_i= (33/5,1,-3)$ and \Mi, $g_i$ for $i=1,2,3$ are the gaugino masses
and gauge couplings (in the standard normalization with $g_1= \sqrt{5/3}\, g_Y$).  This immediately implies
\begin{equation}
 \frac{\Mi(Q)}{g^2_i(Q)} = \mbox{constant} \equiv \frac{\Mi(Q_{in})}{g^2_i(Q_{in})}\label{eq:MoverG2}
\end{equation}
where $Q_{in}$  is an arbitrary initial scale, either the high scale for top-down or the low scale for
bottom-up evolution.  
More explicitly, the one-loop RGE solutions for the gauge couplings lead to 
\begin{eqnarray}
\mathrm{Z}_i  & = & \left[1+\frac{B_i}{4\pi}\alpha_i(Q_{in})\,
\ln(\mathrm{Q}^2/\mathrm{Q_{in}}^2)\right]^{-1} \nonumber \\
\mathrm{M}_i(Q)           & = & \mathrm{Z}_i \mathrm{M}_i(Q_{in}) 
\end{eqnarray}
where $\alpha_i\equiv g^2_i/(4\pi)$. 

The structure of the RGE thus shows that the gaugino sector 
is essentially determined by gauge couplings, and to some extent by the Yukawa couplings which
only 
enter at the two-loop level. The gaugino parameters are decoupled from the scalar 
sector. 
In contrast, the soft parameters in the scalar sector are strongly correlated due to the RGE: 
most of the soft scalar masses depend not only on their value at the initial scale, but also on other
 scalar masses as well as gaugino mass and trilinear coupling parameters. 
An approximation of those
solutions for all possible scalar masses can be written e.g. in terms of high scale universal 
$m_0, m_{1/2}, A_0$ parameters, but 
these are not very useful and can even be misleading.
The RGE for the
slepton and squark parameters of the first two generations, \MselR, \MsqL, \MsqR, $\dots$,  
depend essentially on the gauge couplings, gaugino
masses and \theTrace\ at one-loop~\cite{RGE2}. In fact, 
\begin{equation}
\frac{d}{dt} \MselR^2 = \frac{1}{2\pi} \frac{3}{5}\alpha_1 \left( \theTrace -4 M^2_1\right)
\label{Mser}
\end{equation}
\begin{multline}
\frac{d}{dt} \MsqL^2 = \\
\frac{1}{2\pi} \left( \frac{\alpha_1}{10} \theTrace -\frac{\alpha_1}{15} M^2_1
-3 \alpha_2 M^2_2 -\frac{16}{3} \alpha_3 M^2_3 \right)
\label{Msql}
\end{multline}
with \theTrace\ defined as: 
\begin{equation}
\begin{split}
\theTrace & = \MHTwo^2-\MHOne^2  \\
&\quad + \MsqOneL^2 - \MselL^2 - 2\MsquR^2+\MsqdR^2+\MselR^2  \\
&\quad + \MsqTwoL^2 - \MsmuL^2 - 2\MsqcR^2+\MsqsR^2+\MsmuR^2  \\
&\quad + \MsqThreeL^2 - \MstauL^2 - 2\MstR^2+\MsbR^2+\MstauR^2 
\label{trace}
\end{split}
\end{equation}
i.e. the sum over all scalar soft terms degrees of freedom weighted by their hypercharge. 
This trace has the property of vanishing at tree-level for any model where
some (even partial) universality relations holds among the soft masses, and moreover 
remains constant at one-loop level when evolved to an arbitrary scale $Q$ (i.e. 
$\frac{d}{dt}[\theTrace_{1-loop}]=0$)\cite{Martin:1993ft}. 
At  two-loop RGE order, it gives roughly a relative correction 
of at most 10\% of the largest scalar mass squared. In SPS1a this squared mass is $\MHTwo^2$.     
If the \theTrace\ is zero at one-loop level, i.e., all scalar parameters are well determined
in SPS1a, a moderate increase of the errors on the parameters is expected after their
evolution to the high scale. 
Due to the relatively large coefficient of the $\alpha_3 \MThree^2$ term in Eq.~\ref{Msql} 
the RGEs are sensitive to \MThree.
  
For the third generation scalar masses, the RGEs are more involved and definitely couple different
scalar species.
Typically for the relevant parameters \MstauL, \MstauR, their respective (one-loop) 
RGE both involve (apart from \theTrace) the parameters: \{\MHOne, \MstauL, \MstauR,
\Atau \}, plus the relevant gauge couplings and gaugino masses. 
Thus for instance  the parameter \MstauR\ of the MSSM at the final (low or high) scale after RGE running 
from the initial scale  
will depend on the value of the initial \MstauL value, as well as
the other parameters above, in a complicated way.

To illustrate the impact of these correlations, the precision of the determination of the MSSM parameters 
will be compared to the precision of the high scale MSSM parameters. 

\section{Collider Data}
\label{sec:data}

The SPS1a parameter set leads to moderately heavy squarks and gluinos
in the range of 500~GeV to 600~GeV. The sleptons have masses ranging from
130~GeV to about 200~GeV. The light neutralinos and chargino have masses well
below 200~GeV and their field content is predominantly gaugino, whereas the heavier
states are higgsino. The detailed analyses at the LHC and the ILC can be 
found in Ref.~\cite{Weiglein:2004hn}.

\subsection{LHC and ILC measurements}

At the LHC the SPS1a spectrum can lead to long decay chains, the most prominent being:
\begin{equation}
\tilde{q}_L\rightarrow\chi_2^0 q\rightarrow\tilde{\ell}_R\ell q\rightarrow\ell\ell q \chi^0_1 .
\end{equation}
In this decay chain at least five observables can be determined~\cite{Bachacou:1999zb,Allanach:2000kt}. 
The observables
are endpoints or thresholds of invariant mass combinations among the leptons and
jets. Additional measurements cover essentially the strongly interacting
sector. In SPS1a it is difficult to observe the stop quarks above the supersymmetric
background from sbottom decays leading to the same final state. The use of 
stop sector branching ratios has been explored in~\cite{Bechtle:2009ty,mihoko}.
Stop kinematic edges have been studied for other
benchmark points~\cite{Ball:2007zza,Casadei:2010nf}.
While the corresponding results have not 
been experimentally confirmed, recent progress in fat-jet analysis techniques 
indicates that by the time LHC acquires a sufficient luminosity we should be
able to measure the stop mass as well~\cite{stops}. 

Parts of the electroweak sector, namely three of the four neutralinos, $\chi_1^0$, $\chi_2^0$
and $\chi_4^0$
but not $\chi_3^0$ 
will be observed at the LHC. The absence of the fourth neutralino leads to ambiguities, e.g.,
one could suppose that $\chi_3^0$ has been measured instead of $\chi_4^0$. 
In Ref.~\cite{Lafaye:2007vs} an example of the consequences of 
such a wrong assignment is discussed. In the following
such discrete ambiguities will be left out of the discussion. 

In the slepton sector the first and second generation selectrons and smuons will be measured
as well as the lightest stau.
The expected precision at the LHC for the measurements is listed in 
Table~\ref{tab:LHCobs} for an integrated luminosity of 300~fb$^{-1}$. In the analysis, 
the channels involving leptons have been separated for muons and electrons. 
The systematic error of each channel was unchanged, but the statistical error was increased
to take into account the reduced statistics per observation. This approach should be considered
as the optimistic limit of what can be done at the LHC as additional 
systematics, e.g., due to a more difficult fit of the background, would have to be added. 
Note that even with the increase of the 
statistical error, the systematic (energy scale) error still dominates the experimental error.
Additional
observables such as the use of cross sections times branching ratios have been studied in~\cite{Dreiner:2010gv}.
The masses of the sparticles can be derived from the observables listed in Table~\ref{tab:LHCobs}
with a fit or toy experiments without the use of the underlying theory~\cite{Weiglein:2004hn}.

At the ILC essentially all kinematically accessible sparticles
can be measured. Masses are measured 
either in direct reconstruction at a center--of--mass energy higher than the 
production threshold or via a measurement of the production cross section
as function of the center--of--mass energy at the threshold of (s)particle
production. As the beam energy is well known from the 
accelerator, typically the expected precision of the mass measurements 
is about an order of magnitude better than at the LHC. 
The ILC observables are shown in Table~\ref{tab:ILCobs}.

\begin{table}[htb]
\begin{center}
\begin{tabular}{lc@{$\pm$}c@{$\pm$}c}
\hline\noalign{\smallskip}
particle & $m_{\rm SPS1a}$ value & stat.err. & theo.err.\\
\noalign{\smallskip}\hline\noalign{\smallskip}
$h$  & 108.7 & 0.05 & 2.0 \\
$H$  & 395.34 & 1.5  & 2.0\\
$A$  & 394.9 & 1.5  & 2.0\\
$H+$ & 403.5 & 1.5  & 2.0\\
$\chi_1^0$ &  97.22& 0.05  & 0.5\\
$\chi_2^0$ & 180.44& 1.2   & 0.9\\
$\chi_3^0$ & 355.45& 4.0   & 1.8\\
$\chi_4^0$ & 375.09& 4.0   & 1.9\\
$\chi^\pm_1$ & 179.79 & 0.55 & 0.9\\
$\chi^\pm_2$ & 375.22 & 3.0  & 1.9\\
$\tilde{t}_1$ & 398.93&  2.0  & 4.0\\
$\tilde{e}_L$    & 199.59   & 0.2  & 1.0\\
$\tilde{e}_R$    & 142.68   & 0.05 & 0.7\\
$\tilde{\mu}_L$  & 199.59   & 0.5  & 1.0\\
$\tilde{\mu}_R$  & 142.68   & 0.2  & 0.7\\
$\tilde{\tau}_1$ & 133.36   & 0.3  & 0.7\\
$\tilde{\tau}_2$ & 203.62   & 1.1  & 1.0\\
$\tilde{\nu}_e$  & 183.72   & 1.2  & 0.9\\
\noalign{\smallskip}\hline
\end{tabular}
\end{center} 
\caption[]{Errors for the mass determination in SPS1a, taken  
from~\cite{Weiglein:2004hn}. Shown are the nominal parameter values  
(from SUSPECT~\cite{Djouadi:2002ze}), fixing the electroweak symmetry breaking and renormalization scales at $1$ TeV,
  the error for the ILC alone as well as the theoretical error, all in units of GeV.}
\label{tab:ILCobs}
\end{table}

As the RGE running depends strongly on the top quark Yukawa coupling value and its error, the pole mass of 
the top quark and the strong coupling constant $\alpha_S$ are included as parameters and measurements in the fit,
in addition to the supersymmetric and soft-supersymmetry-breaking parameters.
An error of 1~GeV is assumed when only LHC data is used~\cite{ATLASTDR,Ball:2007zza}. For parameter determinations
involving the ILC, a theoretical error of 0.12~GeV is used with a negligible statistical error~\cite{Djouadi:2007ik}.
For the strong coupling constant a conservative error estimate of 0.001 is used~\cite{Bethke:2009jm}.

\subsection{Theoretical analysis of the neutralino sector}
\label{sec:crudeTheo}

The measurements of the neutralino masses strongly influence the determination of the parameters. 
Most of the qualitative results of the full analysis in the gaugino/higgsino sector parameters can be understood
from a simplified theoretical analysis, which depends solely on the neutralino sector
parameters, neglecting all errors. The neutralino mass matrix at tree level is :

\begin{equation}
\left(
\begin{array}{cccc} {\MOne} & 0 & -m_Z s_W c_\beta & m_Z s_W s_\beta  \\
 0 & { \MTwo} &  m_Z c_W c_\beta & -m_Z c_W s_\beta  \\
 -m_Z s_W c_\beta & m_Z c_W c_\beta & 0 & -{ \mu }\\
  m_Z s_W s_\beta & -m_Z c_W s_\beta & -{ \mu} & 0 
\end{array} \right)\;
\label{mino}
\end{equation}
where $s_W=\sin\theta_W$, $c_W=\cos\theta_W$, $s_\beta=\sin\beta$, $c_\beta=\cos\beta$. 
The dominant radiative corrections to the neutralino masses~\cite{corr1,corr2} are incorporated
in the form of $\MOne \to \MOne+\Delta \MOne$,  $\MTwo\to\MTwo+\Delta \MTwo$, $\mu \to \mu+\Delta \mu$
($\tan\beta$, $m_Z$ and $s_W$ can also be considered as the radiatively corrected values in the 
$\overline{DR}$
scheme). Assuming that the four neutralino masses are measured, 
fixing $\tan\beta$ temporarily, the first approximation is $m_Z\to 0$: in this case, the correspondence between
the mass
eigenvalues and basic parameters of Eq.~\ref{mino} is trivially given as 
\begin{equation}
m_{\chi_i^0} (i=1,..,4) = \MOne, \MTwo, |\mu|, |\mu|\; 
\end{equation} 
with all possible permutations, 
i.e. one obtains a 12-fold degeneracy, corresponding to the six possible permutations among the parameters
\MOne, \MTwo, $|\mu|$ and the ambiguity of the sign of $\mu$.  
Restoring the full $m_Z$ dependence renders the solution more complex but qualitatively similar: 
given that the three neutralinos are measured at the LHC, \MOne, \MTwo\ and $\mu$ are determined 
from the following system of three equations~\cite{inv1,Kneur:2008ur}:
\begin{eqnarray}
& P^2_{ij}+(\mu^2+m_Z^2-\MOne\MTwo +(\MOne+\MTwo)S_{ij}-S_{ij}^2)P_{ij}  \nonumber \\
& +\mu m_Z^2(c_W^2 \MOne+s_W^2 \MTwo) \sin 2\beta-\mu^2 \MOne\MTwo=0\;
\label{b4}
\end{eqnarray}
and
\begin{eqnarray}
&(\MOne+\MTwo-S_{ij})P^2_{ij}+(\mu^2(\MOne+\MTwo) \nonumber \\ 
& +m_Z^2(c_W^2 \MOne+s_W^2 \MTwo-\mu \sin 2\beta))P_{ij}\nonumber \\
& +\mu( m_Z^2(c_W^2 \MOne +s_W^2 \MTwo) \sin 2\beta-\mu \MOne\MTwo)S_{ij} \nonumber \\
& =0 \;
\label{b5}
\end{eqnarray}
with    
    $S_{ij} \equiv m_{\chi_i^0} + m_{\chi_j^0} $, 
     $P_{ij} \equiv m_{\chi_i^0} m_{\chi_j^0} $ 
     for any pair of neutralinos $i,j=1,2,4$~\footnote{These equations are symmetrical
under all neutralino mass permutations.}, and
\begin{equation}      
      \mu^2 = \MOne\MTwo -m^2_Z-\left(P_{124}+S_{124}( \MOne + \MTwo -S_{124})\right)\;
\label{mu3N}
\end{equation}
where $S_{124}\equiv  m_{\chi_1^0}+m_{\chi_2^0}+m_{\chi_4^0}$
and $P_{124} \equiv  m_{\chi_1^0} m_{\chi_2^0} + m_{\chi_1^0} m_{\chi_4^0}+
 m_{\chi_2^0} m_{\chi_4^0}$.

For SPS1a this system can easily be solved numerically to obtain the full set of degenerate solutions
for \MOne, \MTwo, $\mu$, labeled DS1 to DS12 in Table \ref{tab:12fold}. These solutions 
cover all possible
hierarchies among $|\MOne|$, $|\MTwo|$ and $|\mu|$. Due to $m_Z\ne 0$, they 
no longer correspond to simple permutations. The six different hierarchies remain clear and the solutions
for the opposite sign of $\mu$ are not exactly symmetrical. 
In fact the system Eq.~\ref{b4}, Eq.~\ref{b5} and Eq.~\ref{mu3N} gives 12 solutions, not necessarily all real-valued,
for any fixed neutralino mass input. 
Taking different possible (physically irrelevant) 
sign choices for the input neutralino masses exhausts all possible solutions 
of different $\mu$, \MOne\ signs within the six different hierarchies. In the present
study only solutions with positive \MOne\ are considered for simplicity. 
Simple approximate solutions are derived in
Ref.~\cite{Kneur:2008ur} by expanding Eqs.~\ref{b4}-\ref{mu3N} to first order in $m^2_Z$.
The difference with respect to the values in Table~\ref{tab:12fold} is about one percent.

\begin{table}[htb]
\begin{center}
\begin{tabular}{cccccc}
\hline
solution & \MOne\ & \MTwo\ & $\mu$ & $m_{\chi_{\mathrm{pred}}^0}$  \\
\hline 
DS1& 97.66  &  187.35 &  -358.43 & 367.7       \\
DS2& 182.44  &  98.54 &  -361.64 & 371.8      \\
DS3& 102.35  &  354.88 &  -184.61 & 195.5     \\
DS4& 368.7  &  120.16 &  -165.49 & 197.0     \\
DS5& 168.0  &  357.44 &  -115.27 & 127.3     \\
DS6& 369.45  &  144.05 &  -77.94 & 55.2      \\
DS7& 100.41  &  196.68 &  349.34 & 355.6    \\
DS8& 184.73  &  106.47 &  354.69 & 361.5     \\
DS9& 109.26  &  350.25 &  185.84 & 193.2     \\
DS10& 367.13  &  140.59 &  170.86 & 215.9      \\
DS11& 163.59  &  354.52 &  126.55 & 134.6       \\
DS12& 368.88  &  136.13 &  83.44 & 46.7         \\
\hline
\end{tabular}
\end{center}
\caption[]{The 12 degenerate solutions found in the theoretical analysis of the neutralino sector. All values are in GeV.}
\label{tab:12fold}
\end{table}
All 12 degenerate solutions are compatible with a consistent radiative electroweak
symmetry breaking $|\mu|$ solution, provided that the values of the other parameters
in the Higgs sector are calculated consistently from this value of $\mu$. 

The values of the remaining neutralino mass ($\chi_{\mathrm{pred}}^0$), uniquely predicted for any 
of the 12 solutions, are also given in Table~\ref{tab:12fold}.
As expected, $\chi_{\mathrm{pred}}^0$ strictly speaking distinguishes the 12 solutions,
but it is not measured at the LHC. 
In eight of the solutions the $\chi_{\mathrm{pred}}^0$ is almost as heavy as the $\chi_4^0$, but
not degenerate in mass due to $m_Z\ne 0$. 
In the other four solutions DS5, DS6, DS11 and DS12
this neutralino becomes the LSP or next-to-lightest supersymmetric particle.

\subsection{Errors}

The measurement of unification in the supersymmetric sector
relies not only on a precise estimation of the experimental 
error, but also on a rigorous treatment of the theoretical error.
The theoretical error is 0.5\% for the masses of 
colorless particles, the neutralinos, charginos and sleptons.
The error on the gluino and squark mass predictions is taken 
to be 1\%. The errors reflect the difference between spectrum 
generators calculating the observables with the same precision but using 
different methods, as well as (to some extent) the renormalization scale dependence as a measure of  
not yet calculated higher order terms. For SPS1a, performing with SuSpect a rather
large variation of the renormalization scale, from 200 GeV to 1 TeV, gives variations of the Higgs 
and sparticle masses which are comfortably below the quoted uncertainties in Table \ref{tab:ILCobs}.
To illustrate that also the difference between spectrum generators is covered will be illustrated 
by using SoftSUSY instead of SuSpect. 
At the LHC the masses are not measured directly,
the theory errors are considered to be uncorrelated and propagated to the observables.
For the observables of SPS1a this is a conservative choice as positively correlated
theory errors on masses will lead to smaller errors by a factor~2-3. 

The expected precision of the measurements is shown in 
Table~\ref{tab:LHCobs}. The last column in this table is different
from the one shown in Ref.~\cite{Lafaye:2007vs} as the theoretical errors
of the MSSM predictions are shown here whereas in Ref.~\cite{Lafaye:2007vs}
the errors on the mSUGRA predictions are shown.

The experimental systematic errors at the LHC such as the lepton
energy scale are considered to be 99\% correlated to assure that the correlation matrix can be
inverted even for negligible statistical errors. For the treatment of the 
theoretical error, the RFit scheme~\cite{rfit} is largely followed.
Given a set of measurements $\vec d$ and 
a general correlation matrix $C$

\begin{alignat}{7}
\chi^2     &= {\vec{\chi}_d}^T \; C^{-1} \; \vec{\chi}_d  \notag \\ 
|\chi_{d,i}| &=
  \begin{cases}
  0  
          &|d_i-\bar{d}_i | <   \sigma^{\text{(theo)}}_i \\
  \frac{ |d_i-\bar{d}_i | - \sigma^{\text{(theo)}}_i}{ \sigma^{\text{(exp)}}_i}
  \qquad  &|d_i-\bar{d}_i | >   \sigma^{\text{(theo)}}_i \; ,
  \end{cases}
\label{eq:flat_errors}
\end{alignat}

where $\bar{d}_i$ is the $i$-th data point predicted by the model
parameters and $d_i$ the measurement. 
The contribution to the $\chi^2$ of a given measurement is 
zero within one unit of the theoretical error. This ensures that
no particular parameter value is privileged within the theoretical
error range. This type of behavior is appropriate for theoretical
errors as, e.g., higher order corrections will necessarily lead to
a shift of the prediction within the region of the theoretical error.
The shift, contrary to experimental errors,
has no reason to be distributed like a Gaussian. 
Outside
of the theoretical range the experimental error is used. 

\section{Parameter determination at the electroweak scale}
\label{sec:paramEWscale}

To find the true parameter set from a set of measurements which have
statistical errors, theoretical errors and correlations in a highly correlated
system, the SFitter framework has been developed. SFitter provides several
algorithms to search for the $\chi^2$ minima (or log-likelihood maxima):
weighted Markov chains~\cite{Lafaye:2007vs}, a Grid approach and a gradient fit (MINUIT).   

SoftSUSY~\cite{Allanach:2001kg} and SUSPECT~\cite{Djouadi:2002ze} provide predictions for 
the mass spectrum corresponding to the chosen MSSM parameters. 
The mass spectrum is turned into a 
set of expected measurements, which are used as input to the MSSM parameter fit.
The prediction of the relic density is calculated with micrOMEGAS~\cite{Belanger:2008sj,Belanger:2006is}.
Unless stated otherwise, SUSPECT is used in the following.

Using the MINOS algorithm from the MINUIT suite of tools, the parameters can be
determined in a single fit together with their errors.
However, the analytical propagation of the errors as function
of the scale is quite tedious and not always possible, therefore
toy experiments are used. Typically 5000 (toy) datasets are generated,
where the expected measurements are smeared according to their experimental and 
theoretical error, including correlations. The determination of the
parameters is performed for each one of the datasets. 
SoftSUSY and SUSPECT provide the RGE running of the fitted parameters between the EW and the GUT scale. 
At any given scale, the width of the parameter distributions, either the RMS (Root-Mean-Square) or the sigma of 
a Gaussian fit, is the error on the parameter, the mean is the central
value of the parameter. From here on a parameter set is defined to be the best-fit result 
for a given toy dataset and the RMS is quoted as error. 

\subsection{Parameter determination from LHC observables}
\label{sec:LHCfit}

The number of observables at the LHC is smaller than the number of supersymmetric parameters
to be determined. Therefore two parameters, for which the LHC has 
small or no sensitivity, are fixed. Fixing the parameters to the true SPS1a values is 
a solution which can be justified a posteriori when grand unification has been proved. 
In this study we have taken a more conservative approach of deliberately fixing two trilinear
couplings, \Atau\ and \Abottom\ to the central value of the allowed parameter range, i.e., to 0~GeV.
Of course, 
by fixing parameters in a correlated system, the errors on other parameters are reduced artificially. 

The fixing of the two parameters is de facto a shift of 250~GeV in \Atau\ and of 750~GeV in
\Abottom. 
Using the nominal values of the other parameters and the true dataset without smearing but 
with  theory errors, the $\chi^2$ remains at zero. However, performing the same calculation
without the theory errors, the $\chi^2$ is~0.8. Thus indeed the two fixed parameters have only a small
impact on the prediction of the LHC observables.
The two major contributions to the $\chi^2$ are the edges involving the sbottom masses and 
the gluino mass. 
As \MThree\ as well as the squark mass parameters of the third generation are free, these can compensate
the shift in the prediction introduced by fixing \Abottom. Since the sbottom masses are also 
used in measurements which involve neutralinos and right--handed sleptons of the first generation, other
parameters such as \tanb, \MOne\ and \MTwo\ are also affected. 
The magnitude of the shift will be discussed later.
Suffice it to say at this stage that the effective shift of the parameters depends also on the effective
correlation of the measurements, either via the explicit correlations (energy scale) or that
introduced by the flat theory errors. The central values of the fits with and without 
theory errors are therefore not expected to be the same. The latter expectation is verified by defining the
theory errors as Gaussian systematic errors and compared to the case where no theory errors
are used. The results for \MOne\ and \MTwo\ differed by about 1~GeV in the two cases.

\begin{table*}[htb]
\begin{center}
\begin{tabular}{lr@{$\pm$}lr@{$\pm$}lr@{$\pm$}lr@{$\pm$}lr@{$\pm$}lr@{$\pm$}lr@{$\pm$}lr@{$\pm$}l}
\hline\noalign{\smallskip}
     & \multicolumn{2}{c}{DS1}     & 
       \multicolumn{2}{c}{DS2}     &
       \multicolumn{2}{c}{DS3}     & 
       \multicolumn{2}{c}{DS4}     &
       \multicolumn{2}{c}{DS7}     & 
       \multicolumn{2}{c}{DS8}     &
       \multicolumn{2}{c}{DS9}     &
       \multicolumn{2}{c}{DS10}     \\
\noalign{\smallskip}\hline\noalign{\smallskip}
\tanb                         &      12.3 & 5.6  &    12.4 & 5.0  &   14.9 & 9.8   &    8.9 & 5.9  & 13.8  & 7.5 &  12.6 & 7.9 &   19.2 & 14.3 &   23.0 & 15.6  \\
\MOne                   &     102.7 & 7.1  &   189.5 & 6.2  &  107.2 & 9.2   &  383.2 & 9.1  & 105.0 & 6.9 & 191.7 & 6.6 &  116.3 & 7.5  & 380.9  & 9.3 \\
\MTwo                   &     185.5 & 7.0  &    96.  & 6.4  &  356.9 & 8.7   &  114.2 & 10.7 & 194.7 & 7.3 & 105.5 & 7.3 &  354.0 & 8.2  & 137.2  & 9.1 \\
$\mu$                   &    -362.7 & 7.8  &  -364.7 & 6.8  & -186.0 & 8.5   & -167.0 & 9.6  & 353.0 & 7.7 & 357.1 & 8.3 &  188.9 & 7.1  & 172.8  & 8.7\\
$\Delta\chi^2_{\mathrm{ILC}}$ & \multicolumn{2}{c}{73}    &	     
                                \multicolumn{2}{c}{22000} &	     
                                \multicolumn{2}{c}{1700}  &	     
                                \multicolumn{2}{c}{25000} &	     
                                \multicolumn{2}{c}{0.4}   &	     
                                \multicolumn{2}{c}{22000} &	     
                                \multicolumn{2}{c}{2000}  &	     
                                \multicolumn{2}{c}{24000} \\	     
								     
ILC                          & \multicolumn{2}{c}{$\tilde{\tau}_1$}  &
                                \multicolumn{2}{c}{$\chi^\pm_1$}      &
                                \multicolumn{2}{c}{$\chi^0_3$}        &
                                \multicolumn{2}{c}{$\chi^\pm_1$}      &
                                \multicolumn{2}{c}{$\tilde{\tau}_1$}  &
                                \multicolumn{2}{c}{$\chi^\pm_1$}      &
                                \multicolumn{2}{c}{$\chi^0_3$}        &
                                \multicolumn{2}{c}{$\chi^\pm_1$}   \\
$\Omega$h$^2$                 & \multicolumn{2}{c}{$0.17\pm0.07$}  &
                                \multicolumn{2}{c}{$(4\pm 2)\cdot 10^{-4}$}      &
                                \multicolumn{2}{c}{$0.14\pm 0.08$}        &
                                \multicolumn{2}{c}{$(8\pm 4)\cdot 10^{-4}$}      &
                                \multicolumn{2}{c}{$0.16\pm 0.07$}  &
                                \multicolumn{2}{c}{$(4\pm 3)\cdot 10^{-4}$}      &
                                \multicolumn{2}{c}{$0.11\pm 0.06$}        &
                                \multicolumn{2}{c}{$(9\pm 4)\cdot 10^{-4}$}   \\
\noalign{\smallskip}\hline
\end{tabular}
\end{center}
\caption{The result of the parameter determination in the gaugino-higgsino sector is shown for the 
eight degenerate solutions at the LHC, including theory errors. DS7 is the true solution (SPS1a).
The increase of the $\chi^2$ when adding the ILC measurements is shown together with the dominant source
of the increase. The last line is the $\Omega$h$^2$ prediction from the LHC measurements. All masses are in GeV.}
\label{tab:mssm_lhc}
\end{table*}

Using Markov chains in the analysis of the weak-scale parameters of the MSSM,
as pointed out in Ref.~\cite{Lafaye:2007vs}, eight degenerate solutions are observed with 
the LHC data set. They cannot be distinguished from each other via the analysis of the $\chi^2$ of the
best--fit result as they are zero when theoretical errors are included. 
While the higgsino/gaugino sector is violently different and distinct, the other parameters are shifted only slightly between
the eight solutions. 

The characteristics described in section~\ref{sec:crudeTheo} are observed : for each solution
with a positive sign of the higgsino mass parameter $\mu$, there is an approximate
mirror solution for negative $\mu$, as well as the permutations of \MOne, \MTwo\ and $|\mu|$.

Note the absence of the four expected solutions DS5, DS6, DS11 and DS12 corresponding 
to the ``higgsino LSP'' hierarchies ($|\mu| < (\MOne, \MTwo)$) which drastically change the 
neutralino mass hierarchies.
A mass splitting of more than about 40~GeV between the lightest two neutralinos
cannot be achieved in this scenario, but a mass splitting twice as large 
is necessary for the LHC observables: the typical value of the $\chi^2$ is of the order of 10$^6$. 

In Table~\ref{tab:mssm_lhc} the gaugino-higgsino sector
is shown for all eight degenerate solutions. The numbering defined in 
Table~\ref{tab:12fold} has been kept to allow comparisons, DS7 is SPS1a, i.e., the true solution. 
The errors on the parameters are within 20\%. These degenerate solutions are indeed well defined local minima from 
which a simple gradient fit like MINUIT cannot escape. 
The central values in Table~\ref{tab:mssm_lhc} agree well with the theoretical 
analysis, which neglects all errors, typically to better than one 
standard deviation, validating the results of the two analyses.

\subsection{Parameter determination from LHC+ILC observables}

The addition of the ILC measurements 
allows to lift the degeneracy of the LHC ambiguous solutions easily.
Table~\ref{tab:mssm_lhc} shows the increase of the $\chi^2$  
due to the ILC. Theoretical errors
are included and no smearing is performed. 

The second-to-last line of Table~\ref{tab:mssm_lhc} shows the ILC measurement with the largest
contribution to the $\chi^2$.  The inversion of \MOne\ with \MTwo\ in DS8 (and DS2) is excluded
via the chargino mass measurement which is sensitive to the value of \MTwo. In DS9 (and DS3), where $\mu$ and
\MTwo\ are exchanged, the chargino mass measurement is not the most sensitive measurement as the chargino mass matrix
is blind to the interchange of these two parameters. Only the deviation of their values from an exact exchange
leads to some sensitivity (10\% of the log--likelihood increase). Here the third heaviest neutralino,
not measured at the LHC, leads to a clear distinction with respect to the true solution. 
For DS10 (as well as DS4) again the chargino mass measurement, sensitive to the values of \MTwo\ and $\mu$
provides the most stringent distinction from DS7.

The log-likelihood is calculated using the LHC parameter set where the trilinear couplings \Abottom\ and \Atau\
are fixed to zero, therefore an increase of the log-likelihood is also observed for the true parameter 
set (DS7) where the effect of increasing \Atau\ from $-250$~GeV to zero is observed via the 
mixing in the precisely measured $\tilde{\tau}_1$ mass. In DS1, which, to first order, differs
from the true set only in the sign of $\mu$, the $\tilde{\tau}_1$ mass measurement via the 
mixing provides the highest sensitivity. 

\begin{table}[htb]
\begin{center}
\begin{tabular}{lr@{$\pm$}rr@{$\pm$}rr}
\hline\noalign{\smallskip}
                     & \multicolumn{2}{c}{LHC}     & \multicolumn{2}{c}{LHC+ILC} & SPS1a \\
\noalign{\smallskip}\hline\noalign{\smallskip}
\tanb                &      13.8 & 7.4             &      10.7 & 3.1             &     10.0 \\
\MOne                &     105.0 & 6.9             &     103.1 & 0.7            &    103.1 \\
\MTwo                &     194.7 & 7.3             &     193.0 & 1.6             &    192.9 \\
\MThree              &     568.3 & 11.6            &     568.5 & 7.8            &    567.7 \\
\MstauL              &     321.4 & 248             &     192.4 & 4.7             &    193.5 \\
\MstauR              &     164.3 & 120             &     134.9 & 5.7            &    133.4 \\
\MsmuL               &     196.3 & 7.6             &     194.4 & 1.2             &    194.3 \\
\MsmuR               &     138.0 & 7.0             &     135.8 & 0.6            &    135.8 \\
\MselL               &     196.4 & 7.5             &     194.3 & 0.8            &    194.3 \\
\MselR               &     137.9 & 7.1             &     135.8 & 0.6            &    135.8 \\
\MsqThreeL           &     491.4 & 16.2            &     486.2 & 11.1            &    481.1 \\
\MstR                &     483.4   & 232           &     409.6 & 17.1            &    409.4 \\
\MsbR                &     502.6 & 15.3            &     499.1 & 13.1            &    502.7 \\
\MsqL                &     529.6 & 12.1            &     526.4 & 5.3             &    526.4 \\
\MsqR                &     508.9 & 16.4            &     507.8 & 14.4            &    506.8 \\
\Atau                &\multicolumn{2}{c}{fixed 0}  &    -102.9 & 681             &   -249.3 \\
\Atop                &    -394.4   & 353           &    -497.3 & 74            &   -496.8 \\
\Abottom             &\multicolumn{2}{c}{fixed 0}  &    -274.2 & 1830     &   -764.0 \\
$m_A$                &     558.2  & 271.2          &     394.9 & 1.5             &    394.9 \\
$\mu$                &     353.1 & 7.7             &     350.8 & 2.5             &    351.0 \\
\noalign{\smallskip}\hline
\end{tabular}
\end{center}
\caption[]{Results for the general MSSM parameter determination in
  SPS1a using flat theory errors. The
  kinematic endpoint measurements are used for the LHC 
  and the mass measurements for the ILC. The LHC+ILC column
  is the combination of the two measurements sets. Shown are the nominal
  parameter values and the result after fits to the different data
  sets. The MSSM theory errors are used. All masses are in GeV.}
\label{tab:mssm_ilc}
\end{table}

The results of the determination of the parameters are shown in Table~\ref{tab:mssm_ilc}
for the LHC and for the LHC combined with the ILC.
As discussed in the beginning of the section, for the LHC the gaugino masses are shifted slightly by 1-2~GeV
with respect to the nominal value to compensate for the fixing of \Abottom\ and \Atau. Parameters with large errors
also contribute to this shift. However these
shifts are small compared to the errors of typically 7~GeV. 

The difference of the results listed in Table~\ref{tab:mssm_ilc} with respect 
to the previous publication are the following: for the LHC the MSSM errors discussed 
in Section~\ref{sec:data} are used instead of the mSUGRA errors. Additionally the degenerate
solutions for \Atop\ shown in Ref.~\cite{Lafaye:2007vs} are not separated out leading to a larger error
on \Atop\ and $\tan\beta$.
For the combination
of LHC and ILC, the Higgs mass measurement of the ILC is used instead of 
the LHC measurement. These changes are reflected in the significantly smaller errors on the parameters.

\begin{boldmath}
\subsection{Relic density and \theTrace}
\end{boldmath}

Any observable 
sensitive to the neutralino couplings and its actual Wino, Bino and Higgsino content, rather than only the mass,
can help to disentangle the LHC degenerate solutions. 
The relic density $\Omega h^2$ for a neutralino LSP is extremely sensitive and drastically changes e.g. for (\MOne, \MTwo) 
exchanged hierarchies.  But $\Omega h^2$ is less sensitive to the
$(|\mu|, \MOne)$ exchange and even less to the $\mu$ sign. 
While the detailed analysis of the relic density is beyond the scope of this paper, 
note that two distinct populations can be identified among the eight degenerate solutions. The  
SPS1a relic density of 0.19, a factor~1.7 too high with respect to the WMAP measurement
of $0.1109\pm0.0056$~\cite{Larson:2010gs},
is obtained for the solutions where the \MOne\ is the smallest parameter, 
i.e., in DS1, DS3, DS7 and DS9. The lightest neutralino
is essentially a Bino in the bulk. In all other cases the relic 
density is off by three orders of magnitude. 

It is interesting to note that the trace expression defined in Eq.~\ref{trace}
plays an important role in the stabilization of the results. As discussed in section~\ref{sec:RGE}
this trace is zero up to one-loop in models with universality in the scalar sector, rendering
the first two generation sfermion masses very mildly dependent on other scalar terms than
themselves.  The two--loop corrections lead to a non zero value ($\sim -1.3\cdot 10^4$), which
in relative units is a moderate perturbation within the evolution of most of the scalar masses,
except for e.g. \MselR\ where it is a substantial contribution to its RGE, see Eq.~\ref{Mser}.    
To test the impact of this additional
constraint this trace is required to be compatible with its SPS1a value within 10\%.
Technically \theTrace\ is added to the LHC observables as an additional
observable with a Gaussian error of 10\%. 
The RMS of the stau parameters and the $m_A$ is reduced by a factor 5 to 10. 

A consequence of $\theTrace\simeq 0$ is that in the poorly determined 
third generation the requirement on \theTrace\ will prevent large
values of the stau sector parameters, thus reduce the allowed space for these parameters. 
The strong reduction of the error on the poorly determined parameters 
shows the sensitivity of this single constraint. 
From here on the constraint is not used in any of the studies.

\section{Evolution to the high scale}
\label{sec:extrapol}

The determination of unification of the supersymmetric parameters for the
true central values of SPS1a a priori does not need any special treatment. 
However in a real experiment the measured parameters will be shifted from 
the true values within their error. Due to the coupling introduced in the RGE equations, 
some badly measured parameters will strongly affect the convergence, in particular at the LHC. Therefore there are two separate questions 
to be answered which are intimately related. The first one is whether there 
is a unification of the $N$ parameters and second question is what is the value of the
unified parameter and at which scale is the unification observed. 

\subsection{Bottom-up evolution}

The evolution from the low scale to the high scale is performed in the following
way: for each toy dataset the best-fit parameter set is determined at the EW scale. 
The range from 1~TeV (where the parameters are defined) to $3\cdot 10^{17}$~GeV (beyond 
the scale where grand unification is expected) is divided into 1000~logarithmically
equidistant steps. Using SUSPECT~\cite{Djouadi:2002ze} the parameters are evolved from the EW scale 
to the next scale point. For each of the toy experiments, the fundamental 
parameters are then known at 1000 discrete scale points. 

As far as the RGE are concerned, the evolution of the parameters between two scales, 
for one fixed point in the input parameter space, should be independent
of whether the evolution is performed in a top-down or bottom-up manner (apart from negligible numerical integration errors), 
as the RGE is obviously invertible. 

\begin{figure}[htb]
\resizebox{0.5\textwidth}{!}{%
  \includegraphics{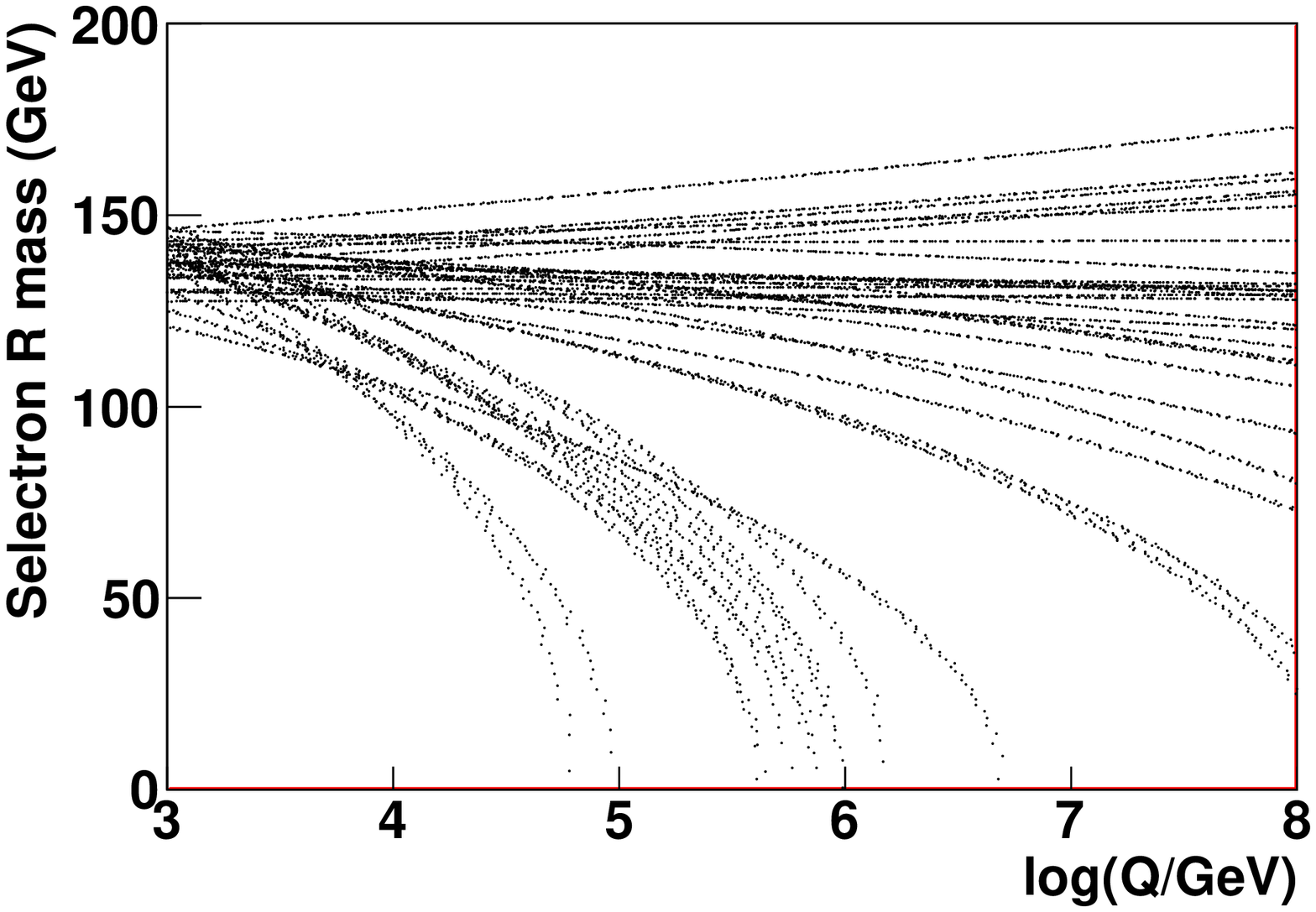}
  \includegraphics{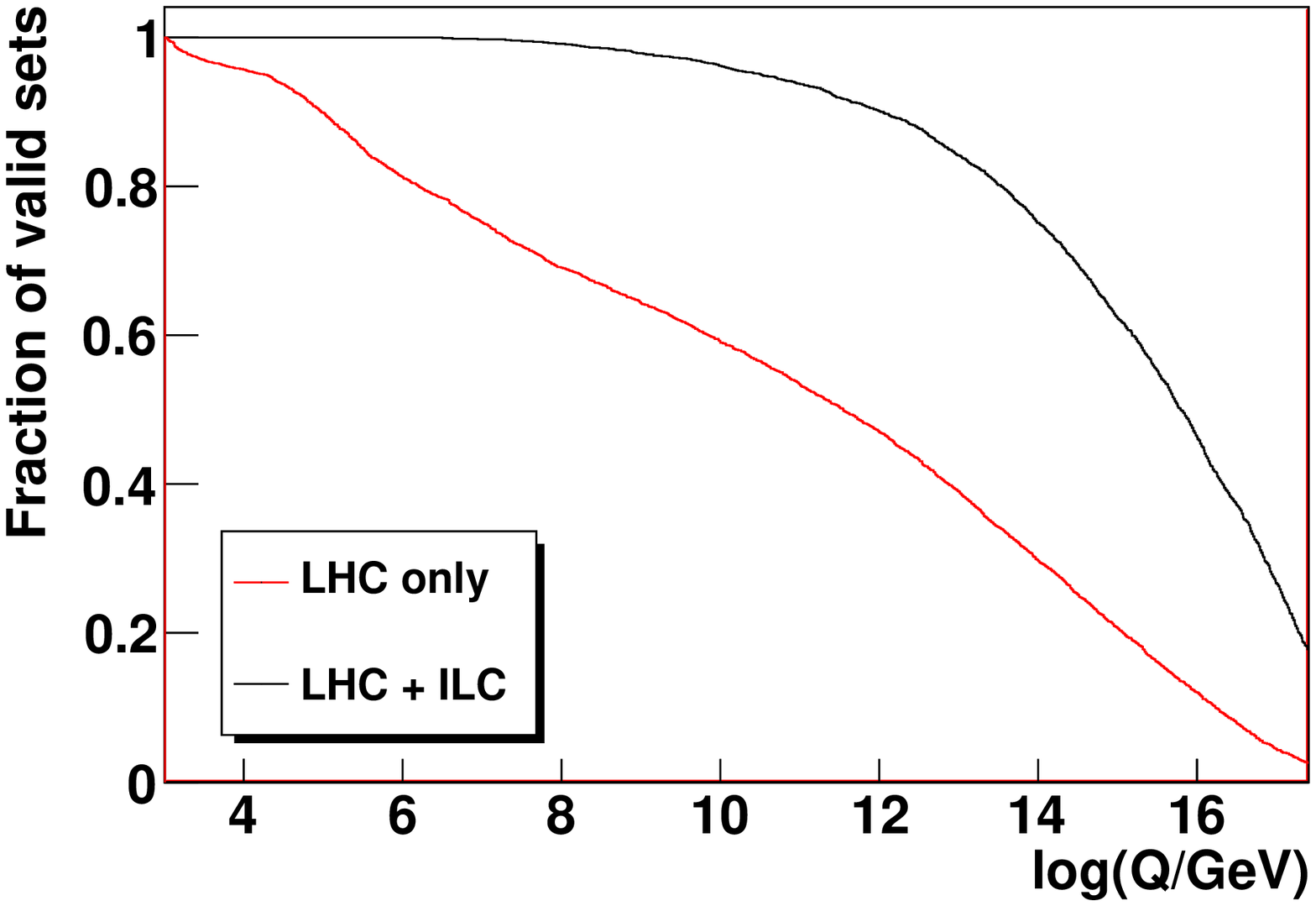}
}
\caption{(left) \MselR\ as function of the logarithm of the scale for LHC measurements.
(right) The fraction of valid non tachyonic parameter sets is shown as function of the 
scale for the LHC and the combination of LHC with the ILC.}
\label{fig:validEvents}       
\end{figure}

However, the errors are amplified strongly as function of the scale, especially in the scalar sector, at least for some parameters. 
This is to a large extent a manifestation of the   
``focus-point'' phenomena in the MSSM~\cite{focuspt}: even if SPS1a does not correspond to what is usually referred as 'focus-point' scenario 
in MSSM (which rather corresponds to much larger $m_0$ values), the focusing behavior is more general,  
i.e. in a large part of the MSSM parameter space the final (low scale) values of some of the scalar parameters 
(in particular \MHTwo\ driving the radiative electroweak symmetry breaking) are not very sensitive to the initial 
(high scale) input choice.     
This means that for the bottom-up direction
even relatively small errors in some of the low scale parameters can result in large errors when evolved at high scale. Typically 
it was shown in Ref.~\cite{Kneur:2008ur} (Table~X in Appendix B) that for relative uncertainties at the low electroweak scale 
of only 1\% in the gluino mass \MThree\ or the up-Higgs doublet mass term \MHTwo\ (letting all other parameters at their
central SPS1a value), the RGE evolution up to the GUT scale amplifies the errors, resulting in relative uncertainties of 20-30\% or even 100\% on some of the final 
high scale soft mass parameters. The parameters entering \theTrace\ are those particularly sensitive to this divergence behavior. 
Therefore, if the initial error is in the few percent range, some of the sfermion masses can become tachyonic well before reaching the
final high scale. 
As an illustration Figure~\ref{fig:validEvents} (left) shows \MselR\ as
function of the logarithm of the scale for all parameter sets. 
\MselR\ is particularly sensitive to the value of \theTrace, as deduced from 
Eq.~\ref{Mser}.
\theTrace\ can deviate substantially from its nominal SPS1a value, e.g. from the largely undetermined
\MstauL\ in Table \ref{tab:mssm_ilc}) for the LHC, and thus drive it to a tachyonic value well before the high scale is
reached.  
A strong non-linear scalar mass dependence enters the RGEs of other scalar masses  
in addition to the \theTrace, such that some tachyonic masses may infect other scalar mass RGE.
All sets which
have at least one tachyonic parameter have to be removed. This necessity is also
confirmed by the analysis of the covariance matrix of the parameters as function
of the scale. If these tachyonic parameter sets are not removed, the covariance matrix
can become singular. 

Figure~\ref{fig:validEvents} (right) shows the percentage of valid non-tachyonic parameter sets.
While at the LHC alone the percentage decreases immediately after the electroweak scale, the addition of the ILC stabilizes
impressively the validity of the sets. For the LHC at 10$^{12}$~GeV only 30\% of the parameter sets
are still valid, whereas with the addition of the ILC 90\% remain. 
At the unification scale for the LHC only 7\% of the parameter sets remain, whereas
for the LHC plus ILC measurements 38\% remain valid, marking a clear improvement over the LHC
alone. Similar results are obtained using SoftSUSY. 

Once a real measurement is available, toy experiments will be defined around the central 
value of the measured data. In the following all confidence level definitions are 
defined with respect to valid, non-tachyonic parameter sets.

Given $N$ parameters for which the grand unification is to be tested,
the following $\chi^2_{avg}$ is to be minimized for every 
scale:
\begin{equation}
\chi^2_{avg} (Q^2) = \sum_{i,j}^N (\Mi - m_{U}) (C_p^{-1})_{ij} (\mathrm{M}_j - m_{U})
\end{equation}
where $C_p$ is the covariance matrix of the parameters and $\mathrm{M}_j$ the $j$-th mass parameter.

The scale where the $\chi^2_{avg}$ is minimal is the best-fit unification scale $Q_{U}$ 
and the parameter $m_{U}$ is the value of the unified parameter. As this procedure
is applied to each dataset, the resulting distribution of all $m_{U}$ and $Q_{U}$
allow to read off the unification scale and unified parameter value as central values
of their distributions and the error as RMS or Gaussian sigma of the distributions. 

A closed formula can be derived for the parameter $m_{U}$~\cite{Schmelling:1994pz}:
\begin{equation}
m_{U}(Q^2) = \left( \sum_{i,j} (C_p^{-1})_{ij} \right)^{-1}\left( \sum_{i,j} (C_p^{-1})_{ij} M_j\right)
\end{equation}

\begin{figure}[htb]
\resizebox{0.5\textwidth}{!}{%
  \includegraphics{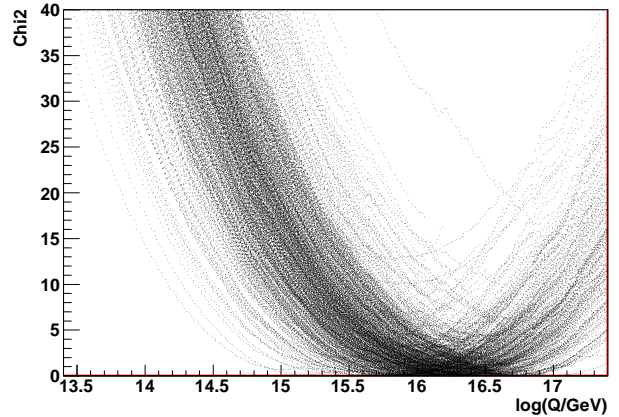}
}
\caption{$\chi^2_{avg}$ of the unification calculation is shown as function of the
scale for DS7. The minimum of the distribution is observed for a scale of about 10$^{16}$~GeV
as expected.}
\label{fig:chi2}       
\end{figure}

However this is not sufficient to claim grand unification as these calculations can also
be performed for non-unifying parameters. To quantify the unification, the absolute
value of $\chi^2_{avg}$ is used. The value is large when the $N$ parameters are not compatible
with a unified one. It is small if the parameters are compatible.
If $\chi^2_{avg}$ is smaller than a cut-off value ($\chi^2_{95}$), the dataset is unified. 
The cut-off is defined so that 95\% of truly unifying
datasets have a $\chi^2_{avg}$ value smaller than $\chi^2_{95}$.
As an example the $\chi^2_{avg}$ for a sample of datasets is shown in
Figure~\ref{fig:chi2} as function of the 
scale (DS7). The minimum at a scale of about 10$^{16}$~GeV is clearly visible. 

\subsection{Evolution of the parameters from LHC observables}

In Table~\ref{tab:mssm_ilc} the result of the determination of the MSSM
parameters is shown in the first column for the LHC. Starting from these values the 
parameter sets are evolved individually to the high scale. 

As noted before, at the LHC an eight-fold ambiguity will be observed.
Therefore the first question is whether the RGE evolution of the eight solutions
will result in similar or different patterns. 

\begin{figure}[htb]
\resizebox{0.5\textwidth}{!}{%
  \includegraphics{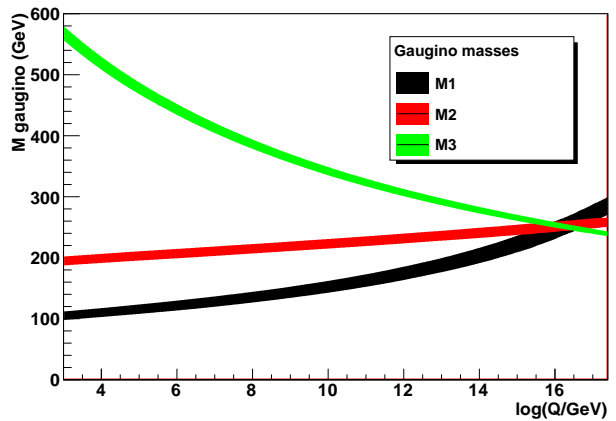}
}
\caption{Evolution of the gaugino mass parameters to the GUT scale for 
DS7 (SPS1a).}
\label{fig:LHCmGauginos}       
\end{figure}

\begin{figure*}[htb]
\resizebox{1.0\textwidth}{!}{%
  \includegraphics{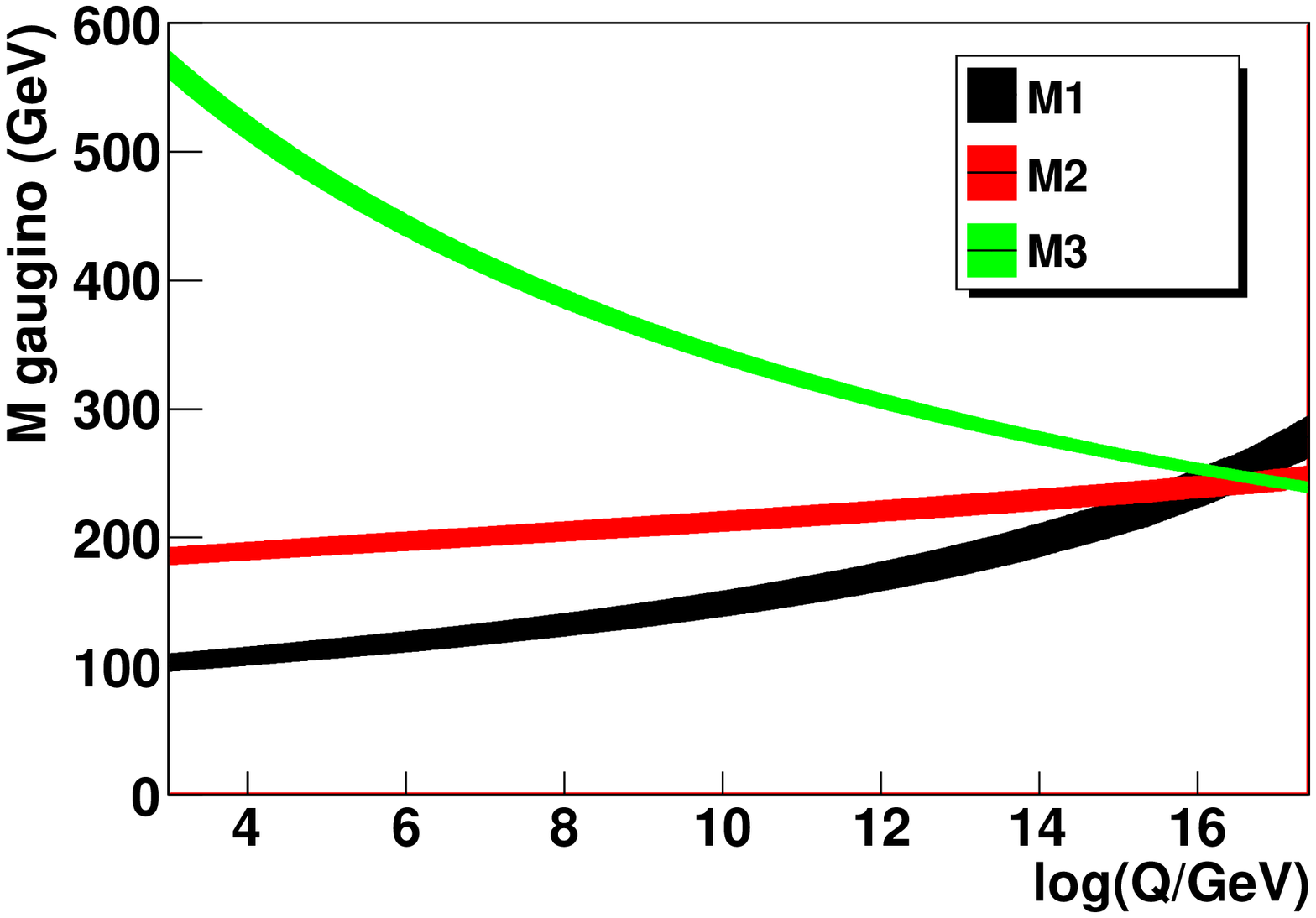}
  \includegraphics{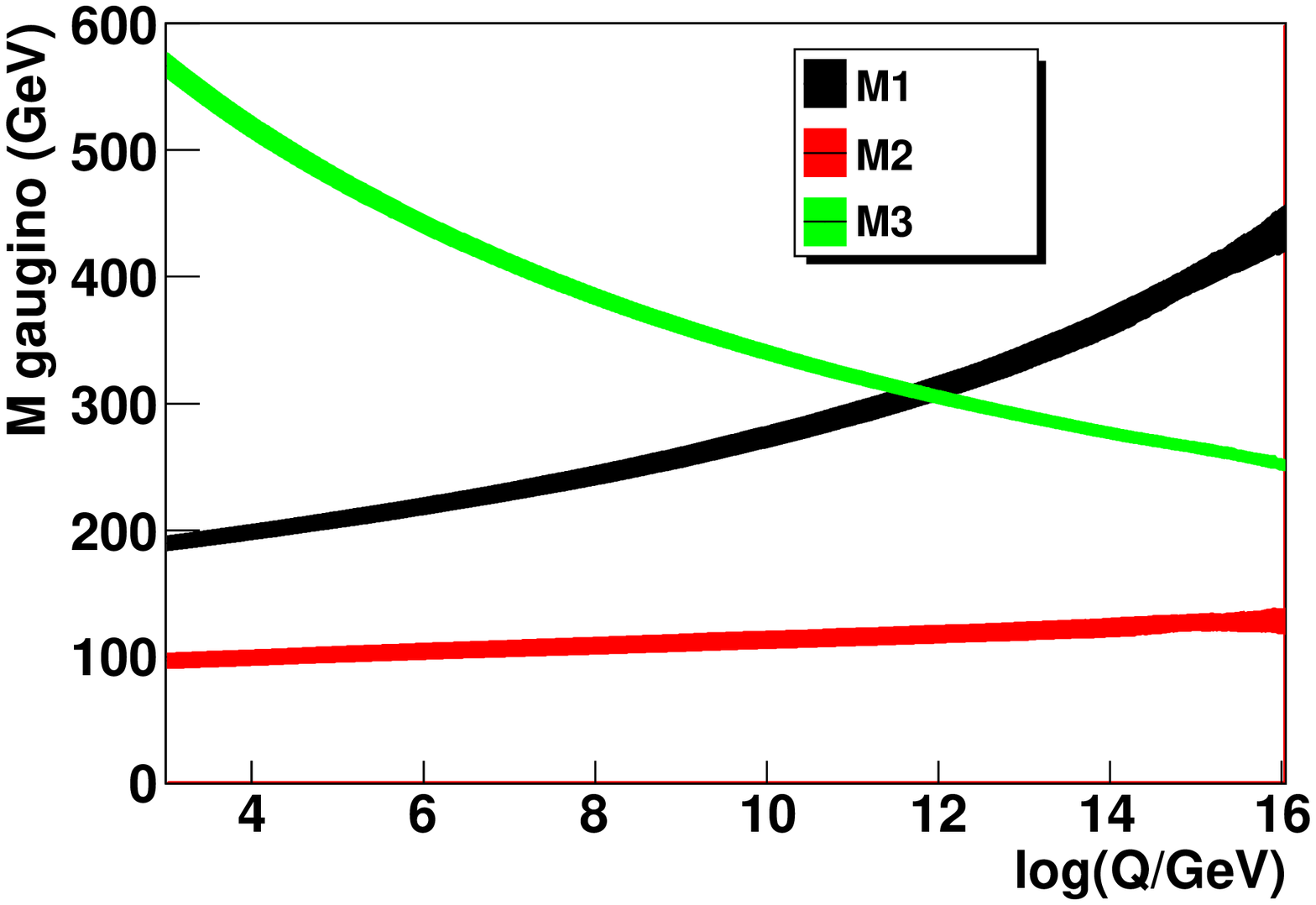}
  \includegraphics{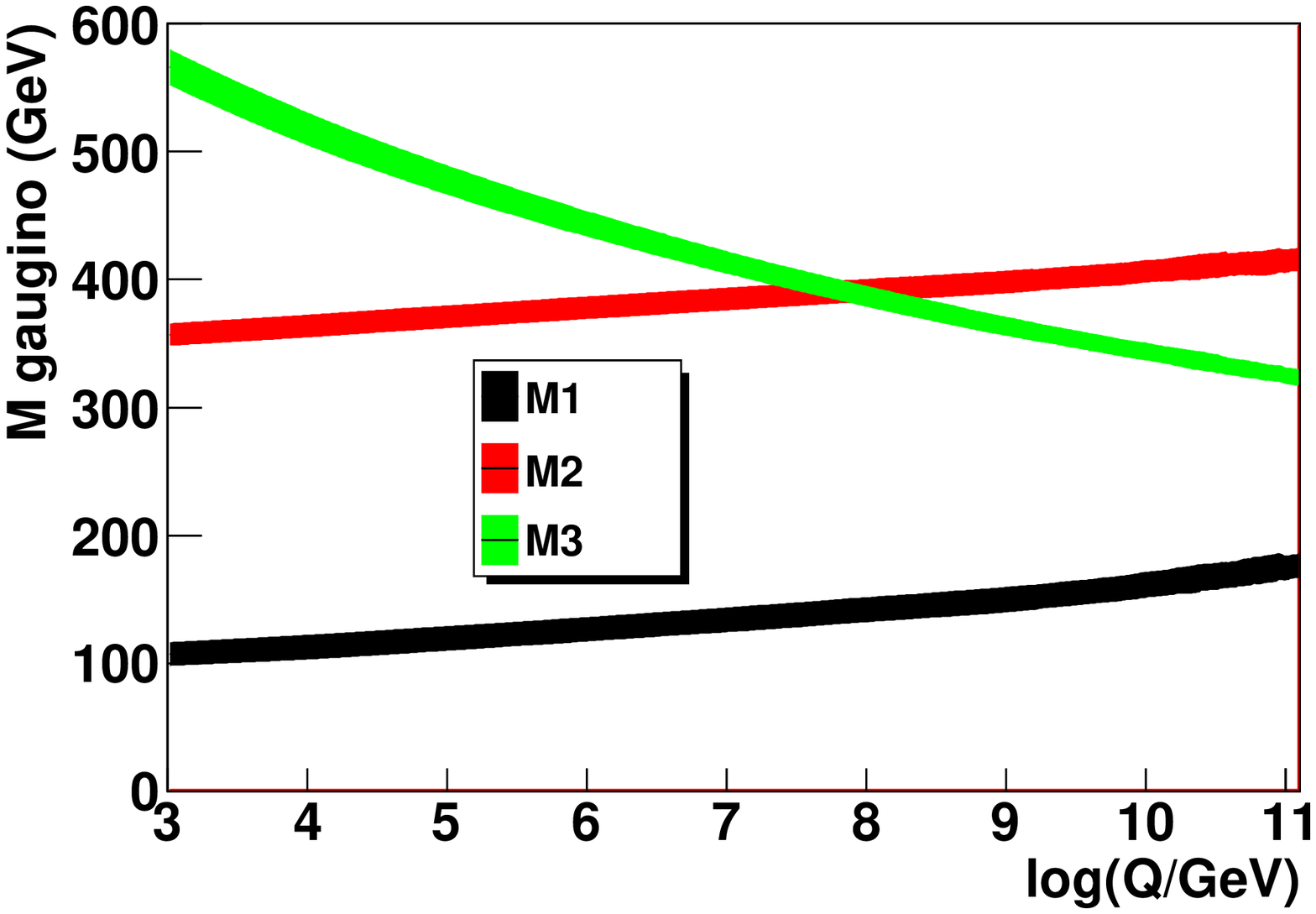}
  \includegraphics{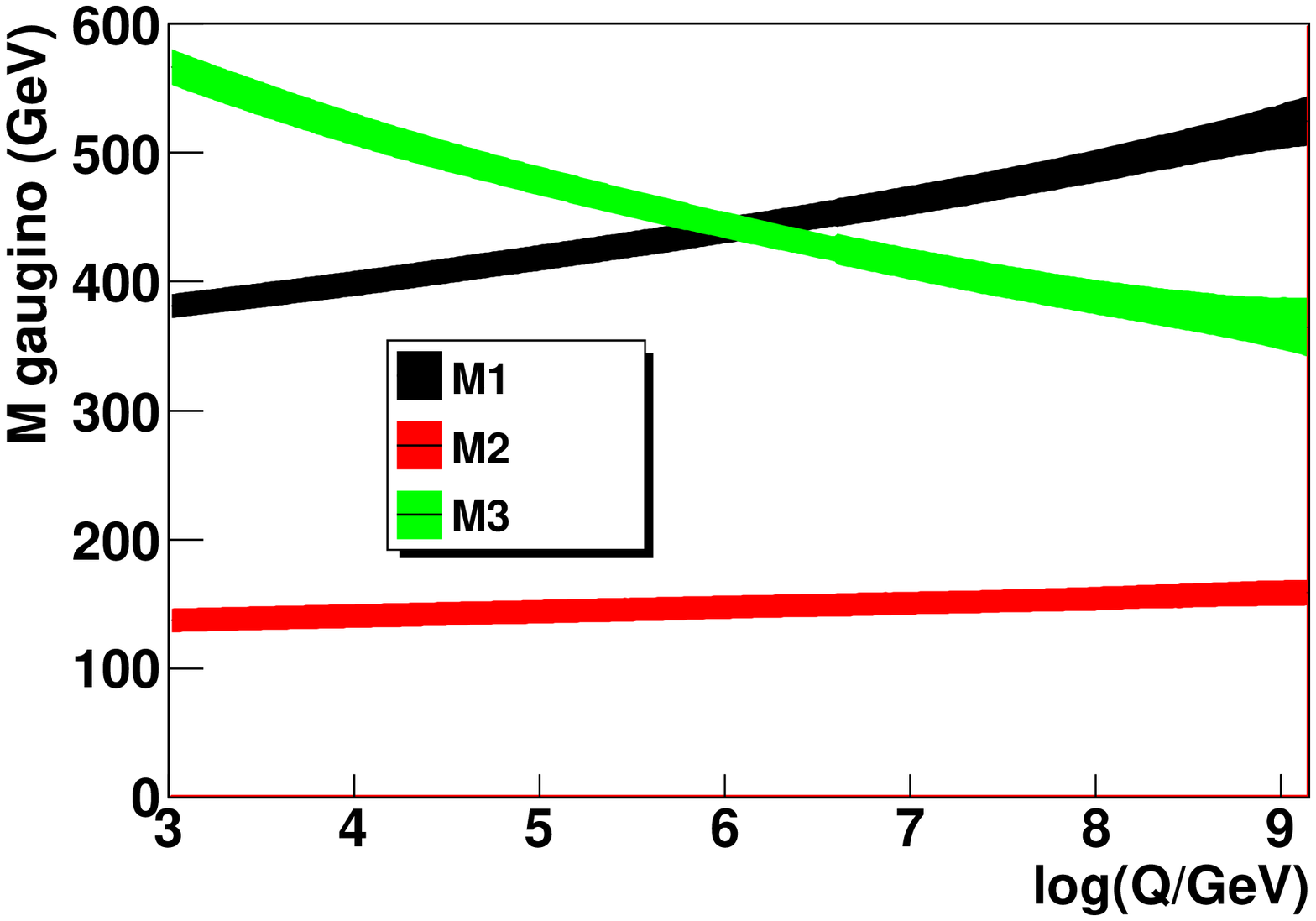}
}
\caption{Evolution of the gaugino mass parameters to the GUT scale for 
the ambiguous DS1, DS2, DS3 and DS10 at the LHC.}
\label{fig:LHCmGauginosFalse}       
\end{figure*}

The evolution of the gaugino mass parameters is 
shown in Figure~\ref{fig:LHCmGauginosFalse} for DS1, DS2, DS3 and DS10 and
in Figure~\ref{fig:LHCmGauginos} for DS7 (SPS1a).
The solutions DS8, DS9 and DS4 show the same pattern as DS2, DS3 and DS10, as expected, as
only the sign of $\mu$ changes. 
In DS2 \MOne\ and \MTwo\ are exchanged with respect to the correct solution. This leads to 
an intersection of \MOne\ and \MThree\ at $10^{12}$~GeV. In DS9 \MThree\ intersects with 
\MTwo\ at about $10^{8}$~GeV, whereas in DS10 \MThree\ and \MOne\ intersect at $10^{6}$~GeV.  

As expected the correct solution unifies the high scale. DS1, qualitatively at least, might
unify. Thus six of the eight ambiguous solutions can be eliminated as candidates for unification.
The difference between DS1 and the true solution being 
only the sign of $\mu$, it is natural that this solution is harder to distinguish from 
the correct one.

A comparison of the number of parameter sets compatible with a unified gaugino mass 
parameter of DS1 and DS7 is therefore necessary to quantify how well one will be able
to distinguish the (non)-unification of these two solutions. 
For the true solution (DS7), at the unification scale determined 
for the gauginos, 95.4\% of the toy experiments 
unify. In DS1 only 38\% are unified. 
Thus the exclusion of
unification for DS1 will indeed be very difficult at the LHC.

The results are in agreement with the expectation from the structure of the RGEs in the gaugino
sector. The absolute value of the gaugino mass measured at the electroweak scale gives the starting
point of the evolution, but the slope is essentially independent of the absolute value, so that
a wrong parameter value at the electroweak scale cannot be compensated. 

As two of the trilinear couplings are fixed at the LHC, no further information on the unification can 
be obtained from these parameters. The same is true for the third generation as the stop sector is
not measured at the LHC. Therefore the study is restricted to the parameters of the first two generations.

\begin{figure}[htb]
\resizebox{0.5\textwidth}{!}{%
  \includegraphics{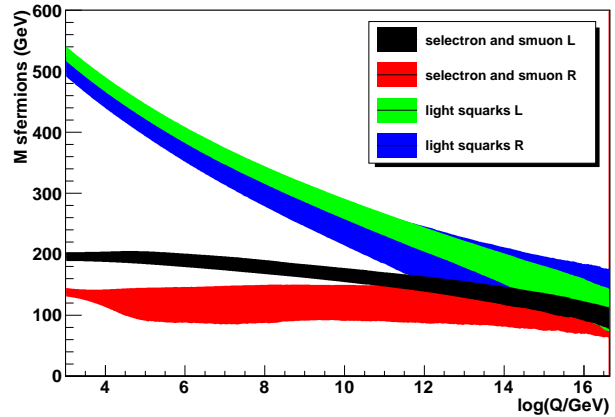}
 }
\caption{Evolution of the first and second generation scalar mass parameters for the true solution (DS7) at the LHC : 
bottom-up evolution of the MSSM.}
\label{fig:LHCmScalars}       
\end{figure}

The bottom-up evolution of the scalar sector for the first two
generations in DS7 is shown in Figure~\ref{fig:LHCmScalars}.
The unification is qualitatively observed at about 10$^{16}$~GeV as expected. 
While the slepton parameters are measured precisely at the electroweak scale, the coupling of the
RGEs leads to a quick degradation of their precision as function of the scale. 
It is obvious
that the scalars will not be able to improve the determination of the unification scale and therefore
will not improve the separation of DS1 and DS7.

\subsubsection{Unification scale and unified parameters}

Given the observation of unification in the gaugino
and scalar sector, the unification scale and the unified parameter can be determined at the LHC.  
Here the study is restricted to the true solution (DS7) without loss of generality, the solution
of DS1 leads to a similar precision. 

\begin{table}[htb]
\begin{center}
\begin{tabular}{lr@{$\pm$}lr@{$\pm$}lr@{$\pm$}lr@{$\pm$}l}
\hline\noalign{\smallskip}
Name                           & \multicolumn{2}{c}{Unified} &
\multicolumn{2}{c}{Unification Scale} & \multicolumn{2}{c}{Parameter at} \\
                               & \multicolumn{2}{c}{Parameter} &
\multicolumn{2}{c}{$[\log(Q/\gev)]$}     &  \multicolumn{2}{c}{1.7 $\cdot 10^{16}$~GeV} \\
\noalign{\smallskip}\hline\noalign{\smallskip}
\mOneHalf                      &  251.9 &  5.9 & 16.23 &  0.29 & 252.3 & 3.2\\
\mZero                         &  98.5 &  10.5 & 16.5  &  0.6 & 100.8 & 4.9\\
\noalign{\smallskip}\hline
\end{tabular}
\end{center}

\caption{\label{tab:LHCunif}Measurement of grand unification with LHC measurements (DS7). All masses are in GeV.}
\end{table}

The results for the gaugino mass parameter \mOneHalf\ as well as
the scalar mass parameter \mZero\ are shown in Table~\ref{tab:LHCunif}.
For the trilinear couplings the calculation is not useful as only 
one parameter is free (\Atop) and the other two are fixed 
at the electroweak scale. 

The most precise determination of the unification scale is obtained in the 
gaugino sector with a measurement of $(1.7\pm 1.1)\cdot 10^{16}$~GeV. 
At the unification scale \mOneHalf\ is measured to about 2\% and is in agreement
with the nominal value of SPS1a (250~GeV). Fixing the scale reduces the error
on the common mass parameter by almost a factor~2.

The common scalar parameter \mZero\ is determined with a precision of about 
10\% in agreement with the nominal value of SPS1a of 100~GeV.
As the scale is measured more precisely in the gaugino sector, combining the two sectors
will not provide an improvement. Alternatively
one can determine \mZero\ at this fixed scale: \mZero\ is measured
to be 101~GeV with an error of 5~GeV, thus the error is reduced by a factor~2,
not including the error on the scale determination.

Thus once the ambiguous solutions for the LHC are discarded, 
the common scalar and gaugino mass parameters can be reconstructed
in a bottom--up approach with a 
precision of 10\% and 2\% respectively. The precision is improved 
to and 5\% and 1\% respectively if the unification scale is fixed. The fixed scale
results can be compared to the mSUGRA parameter determination reported in Ref.~\cite{Lafaye:2007vs}
where a precision of 2\% was reached on the scalar mass and 
roughly 1\% on the on the gaugino mass. Thus in the gaugino sector the results 
of top-down (mSUGRA) and bottom-up (MSSM) agree well, whereas in the scalar 
sector the determination bottom-up is less precise.

\subsubsection{Effect of threshold corrections at the high scale}
\label{sec:threshold}

A further complication arises from threshold corrections of the unknown embedding theory.
The size and the sign of such corrections are model-dependent, 
but in typical SU(5) models these primarily affect~\cite{GUTthresh1,GUTthresh15} 
$\alpha_S(M_{GUT})$ (and $M_{GUT}$ to some extent). This could compensate for the observed 
mismatch, at two-loop RGE in MSSM, in $\alpha_S(M_{GUT})-\alpha_2(M_{GUT})\ne 0$ ($\alpha_2(M_{GUT})\equiv
\alpha_1(M_{GUT}) \equiv g^2_1/(4\pi)$). For a typical minimal SUGRA input, the latter mismatch is a 
few percent and negative (about -3\% in particular for SPS1a). 
In addition, intrinsic corrections to the gaugino masses (i.e. corrections to Eq.~\ref{eq:MoverG2}) have been evaluated to be a few percent in a minimal SU(5) 
model\cite{GUTthresh2}, i.e. roughly of the order of two-loop MSSM corrections, though the former can be much larger in non-minimal GUT models,
e.g. with large representations 
of heavy chiral multiplets. Since specific GUT model corrections are anyway not implemented at present 
in the spectrum calculators, for simplicity a positive shift in $\MThree$ correlated with the $\alpha_S$ one (i.e. preserving Eq.~\ref{eq:MoverG2} 
at one-loop) is assumed. The effect is thus approximated by shifting the measured parameter $\MThree$ by 3\%
for illustration, while possible model-dependent effects on other parameters are neglected. 

The parameter \mOneHalf, including the threshold corrections, 
is shifted by 3.5~GeV and the unification
scale by 0.07, corresponding to a shift of $0.3\cdot 10^{16}$~GeV. 
The absolute values of the shifts have to be compared to the error
of the determination of the common mass (5.9~GeV) and the scale (0.3).
The shift corresponds to a deviation of less than half a standard deviation for the mass, thus
the threshold effects will not play a large role at the LHC,
given the expected precision.

In addition to the study of the central values, it is also interesting to address 
the question whether the threshold effects could lead to the conclusion 
that DS1 unifies and DS7 does not. 
The percentage of the unified parameter sets at the best
scale is a good indicator. Including the threshold corrections, in DS7 the percentage
drops to 87.4\% (from 95.4\%). In DS1, the unification percentage in the gaugino 
sector is 3\% (from 38\%). Thus the threshold corrections applied to DS1 will actually
increase the difference between the true and the wrong solution.
However, if the sign of the $\MThree$ shift is opposite (i.e. if the specific GUT model
is such that those corrections are larger and essentially negative)
 this conclusion would change.

\subsubsection{Evolution with shifted data}
\label{sec:LHCsoftSUSY}

The studies described so far all dealt with datasets which are smeared, but centered around
the true central value. An additional complication will arise with real data as the measured
value of the parameters will be shifted, within the theoretical and experimental errors,
from the true central value. In this case it is still possible to use the toy experiments
but they are performed around the shifted values. 

It is also necessary to verify that the theoretical errors used in the study cover
at least the difference of the predictions from spectrum generators which have similar 
precision. 
The dataset (SPS1a) calculated by SUSPECT
is used, but the MSSM parameters are determined by using SoftSUSY, i.e., SoftSUSY
is used to predict the spectrum and evolve the parameters to the high scale. 
The SPS1a dataset from SUSPECT corresponds to a shifted dataset for SoftSUSY.

The common gaugino mass parameter \mOneHalf\ is determined to be $252.7 \pm 6.4$~GeV
at $\log(Q/\gev)=16.2 \pm 0.3$ with SoftSUSY. 
The difference with respect to the determination using only SUSPECT
is less than about one standard deviation (Table~\ref{tab:LHCunif}). 
In the scalar sector
the common mass is determined to be 92~GeV with an error of 10~GeV. 
The results, both the central values and errors, at the EW scale as well as the GUT scale, are in 
excellent agreement with SUSPECT for the gaugino and scalar sector at the LHC. 

It is also interesting to note that the percentage of toy sets compatible 
with grand unification in the gaugino sector is essentially unchanged: 95\% for DS7,
35\% for DS1.

The results show that as required by the definition of the size of the theoretical 
error at least the difference between different spectrum calculators is covered.
While differences will be observed, depending on which calculator is used,
the difference is small with respect to the expected error on the parameters and the
unification scale.

\subsubsection{High scale MSSM}

To study the difference between bottom-up and top-down evolution, 
5000~toy experiments are used to determine the parameters of the
high scale MSSM. The results are compared 
to the parameters of the MSSM determination after their evolution to the high scale.
Parameter sets are removed as soon as they became tachyonic. Thus they
contribute to the RMS below the scale where they become tachyonic, but not above.

\begin{figure}[htb]
\resizebox{0.5\textwidth}{!}{%
  \includegraphics{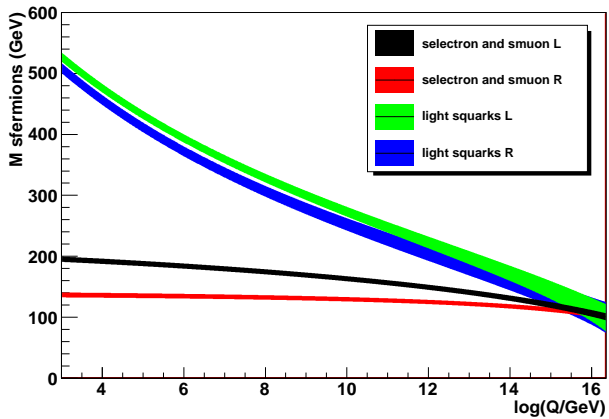}
}
\caption{Evolution of the first and second generation scalar mass parameters for the true solution (DS7) at the LHC : 
top-down evolution for the high scale MSSM.}
\label{fig:LHCmScalarsHighScale}       
\end{figure}

The precision of the parameters is comparable 
in the gaugino sector at the level of 20-30\%. 
In the scalar sector the precision of the high scale MSSM parameters is reduced by
a factor of more than~5. The difference with respect to 
the bottom-up evolution (Figure~\ref{fig:LHCmScalars}) is obvious in Figure~\ref{fig:LHCmScalarsHighScale}.

The apparent contradiction between the naive expectation that top-down should be equal to bottom-up 
and the large differences that are observed, can be traced to several factors:
in contrast to the top-down evolution where 
all parameters are defined at a single
high scale, in the bottom-up case initial conditions for the RGE are more sensitive to the (many) different
physical scales. All threshold corrections are calculated at present
in the one-loop approximation for most sparticles. This induces an increased sensitivity to the
intrinsic errors.   

In the stau sector only one measurement is available at the LHC. This measurement can be used to determine \MstauR. 
In the MSSM the parameter 
\MstauL\ decouples and can take almost any value, even 1~TeV. This is reflected
in the large error in Table~\ref{tab:mssm_ilc}. However, in the high scale MSSM, 
the top-down running introduces interdependencies. 
Keeping all parameters at 
their nominal SPS1a value and moving only \MstauL\ to 200~GeV is not possible:
the selectron and smuon masses are changed by 5~GeV. 
As the lepton-lepton edge, which depends on these slepton masses, is measured precisely at the LHC, 
such a large change is not compatible with the observables. Thus the high scale MSSM restricts the
nominally available parameter space. 

All scalar parameters are correlated non-linearly
through \theTrace\ entering all RGE equations.
In a bottom-up evolution, large errors on the MSSM sfermion parameters 
induce a departure from zero in the boundary (initial) value of \theTrace\. Since the dependence is quadratic, this is strongly amplified in the evolution.
In the top-down evolution as the sfermion parameters are interdependent due to the RGEs and as a consequence
all parameters are well measured, the departure of the \theTrace\ from its SPS1a value is less pronounced and its
effect smaller. 

While the tachyonic parameter sets are removed and play no role at the high scale, they 
contribute to the error on the parameters at all scales below the GUT scale. These parameter sets
are relatively far away from their nominal SPS1a value and thus lead to larger RMS for the parameters
compared to the high scale MSSM where such parameter sets are excluded by construction.

\subsection{Evolution of the parameters from LHC+ILC observables}

In Table~\ref{tab:mssm_ilc} the result of the determination of the MSSM
parameters is shown in the third column for the combination of LHC and ILC.
With the exception of the tri-linear couplings, the parameters are 
measured with excellent precision at the electroweak scale. 
Additionally the eight-fold ambiguity left by the LHC data alone is solved
by the ILC. 

\begin{table}[htb]
\begin{center}
\begin{tabular}{lr@{$\pm$}lr@{$\pm$}lr@{$\pm$}l}
\hline\noalign{\smallskip}
Name                           & \multicolumn{2}{c}{Unified} &
\multicolumn{2}{c}{Unification Scale} & \multicolumn{2}{c}{Parameter at} \\
                               & \multicolumn{2}{c}{Parameter} &
\multicolumn{2}{c}{$[\log(Q/\gev)]$}     &  \multicolumn{2}{c}{2.33 $\cdot 10^{16}$~GeV} \\
\noalign{\smallskip}\hline\noalign{\smallskip}
\mOneHalf                      &  249.5     & 1.8 & 16.37 & 0.05 & 249.6  & 1.5 \\
$\mZero^{1/2\mathrm{Gen}}$     &   98.2     & 10.7  & 16.5 & 0.7 & 100.4  & 2.5 \\
$\mZero^{3\mathrm{Gen}}$       &  117.1     & 27  & 15.4  & 1.3 & 103.1  & 25\\
\mZero                         &  105.3     & 9.1 & 15.9  & 0.6 &  99.4  & 2.0 \\
\AZero                         &  -164      & 182 & 14.8  & 4.5 & -133.8  & 207 \\
\noalign{\smallskip}\hline
\end{tabular}
\end{center}
\caption{The results for the measurement of the common parameters and unification scale with 
LHC+ILC measurements in the bottom-up approach are shown. All masses are in GeV.}
\label{tab:LHCILCunif}
\end{table}

\begin{figure}[htb]
\resizebox{0.5\textwidth}{!}{%
  \includegraphics{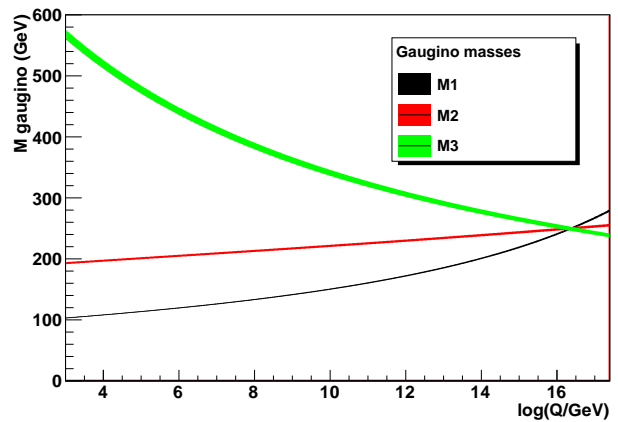}
}
\caption{The evolution of the gaugino masses is shown
for the combined results from LHC+ILC.}
\label{fig:LHCILCmGauginos}       
\end{figure}

The results of the evolution of the parameters measured at the 
electroweak scale as well as the unification are shown in Table~\ref{tab:LHCILCunif}.
The evolution of the three gaugino mass parameters is shown in
Figure~\ref{fig:LHCILCmGauginos}.

The common gaugino parameter \mOneHalf\ is determined with a precision 
of 1.8~GeV at a grand unification scale of $(2.33 \pm 0.28)\cdot 10^{16}$~GeV in 
agreement with the SPS1a parameter set definition. Note that while 
the logarithm of the unification scale is determined with a precision of 0.3\%,
due to the proper error propagation, the scale in GeV is only determined with
10\% precision. The error with respect to LHC alone is reduced by a factor~3.

Fixing the unification scale to the central value determined by the gauginos,
the error on \mOneHalf\ would be reduced by about 0.2~GeV. It is also 
instructive to analyze the individual contributions of the three parameters to the determination
of the unified mass parameter. At the unification scale \MOne, \MTwo\ and \MThree\ are measured
with a precision of 1.6~GeV, 2.0~GeV and 3.3~GeV respectively. Thus the error
on \mOneHalf\ is essentially equal to the precision of \MOne. The naive combination of the 
three parameter errors, i.e., ignoring correlations, would lead to an error of 1.1~GeV, almost 30\% better 
than that obtained. While \MThree\ is uncorrelated with the other two parameters, \MOne\ and 
\MTwo\ are almost 100\% positively correlated. Thus combining the latter two will not increase the precision.
\MThree\ on the other hand, while not correlated, is less precisely measured and therefore the combined error
is decreased only by a small amount in the combination as the error ($\sigma$) on the 
combination of two uncorrelated measurements ($\sigma_1$, $\sigma_2$) reads:
\begin{equation}
\sigma = 1/\sqrt{1/\sigma_1^2 + 1/\sigma_2^2} 
\end{equation}

\begin{figure}[htb]
\resizebox{0.5\textwidth}{!}{%
  \includegraphics{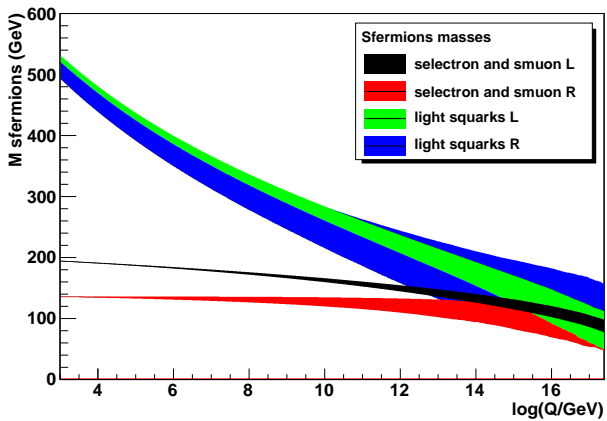}
}
\caption{The evolution of the scalar masses of the first generation is
is shown for the combined results from LHC+ILC.}
\label{fig:LHCILCmScalars}       
\end{figure}
Due to the increased precision of LHC+ILC, 
the unification of the scalar mass parameters can be separated into the first two generations
and the 3rd generation. 
The evolution of the scalar masses of the first generation 
is shown in Figure~\ref{fig:LHCILCmScalars}.
As shown in Table~\ref{tab:LHCILCunif},
$\mZero^{1/2\mathrm{Gen}}$ can be determined with a precision of about 10\% in 
agreement with the nominal value (100~GeV) of SPS1a. The determination of 
the logarithm of unification scale is less precise than the precision in the gaugino
sector by an order of magnitude.
Using the unification scale defined by the gaugino measurement, 
the error is reduced to 2.5~GeV. 
It is interesting to note that in the scalar case the naive combination of the
parameters neglecting the covariance matrix would lead to an error on $m_{0}^{1/2Gen}$
of 8.2~GeV, thus greater than the correct value. This is due to the large negative 
correlations among the parameters which reduce the total error.

\begin{figure}[htb]
\resizebox{0.5\textwidth}{!}{%
  \includegraphics{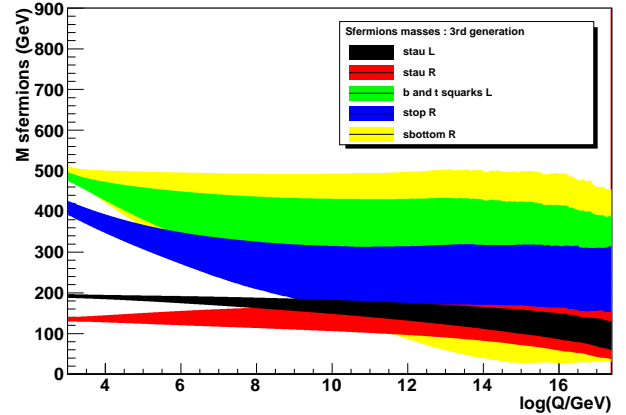}
}
\caption{The evolution of the scalar parameters of the third generation
is shown for the combined results from LHC+ILC as function
of the logarithm of the scale.}
\label{fig:LHCILC3rdGen}       
\end{figure}

For the 3rd generation the unification procedure leads to a less precise
determination (27~GeV) of the common scalar parameter as can be immediately
inferred from Figure~\ref{fig:LHCILC3rdGen} as well as Table~\ref{tab:LHCILCunif}. 
The improvement by fixing the unification is rather small.
The larger errors with respect to the first two generations are due to two sources.
The parameters of the third generations are less precisely measured than 
those of the first two generations. Additionally,  
as discussed in Section~\ref{sec:RGE} the sfermion mass terms 
in the RGE have a stronger inter-dependence.
Nevertheless the reconstructed
unification scale is in agreement with the SPS1a parameter set.

If instead of separating the first and second generation from the third generation (and the two Higgs 
parameters), all parameters relating to scalars are combined, the error on the determination
of the unification scale as well the parameter is decreased as shown in the second to last row
of Table~\ref{tab:LHCILCunif}. 
Using the central value of the unification scale determined by the gaugino sector, the error 
on \mZero\ is reduced to 2~GeV. This error is smaller than the naive combination ignoring correlations
by a factor~3, showing the necessity of a proper treatment of the errors and correlations.

It is interesting to note that the error here is identical 
to the error from the LHC alone. This might seem surprising at first sight as the slepton 
sector is measured experimentally an order of magnitude more precisely at the ILC than at the LHC.
The reason for this (superficial) lack of impact lies again in the structure of the RGEs. 
The error on the sleptons, as shown in Figure~\ref{fig:LHCILCmScalars}, increases strongly 
as function of the scale due to the coupling with the less precisely determined squark sector. 
Thus at the unification scale, the slepton precision of the ILC is diluted by the LHC squark precision.
Additionally while for the combination of LHC and ILC no MSSM parameters are fixed, for the LHC
the trilinear couplings of the sbottoms and staus are fixed. Fixing these leads to contraction
of the allowed parameter space and therefore an artificial decrease of the scalar mass error
at the LHC. 

\begin{figure}[htb]
\resizebox{0.5\textwidth}{!}{%
  \includegraphics{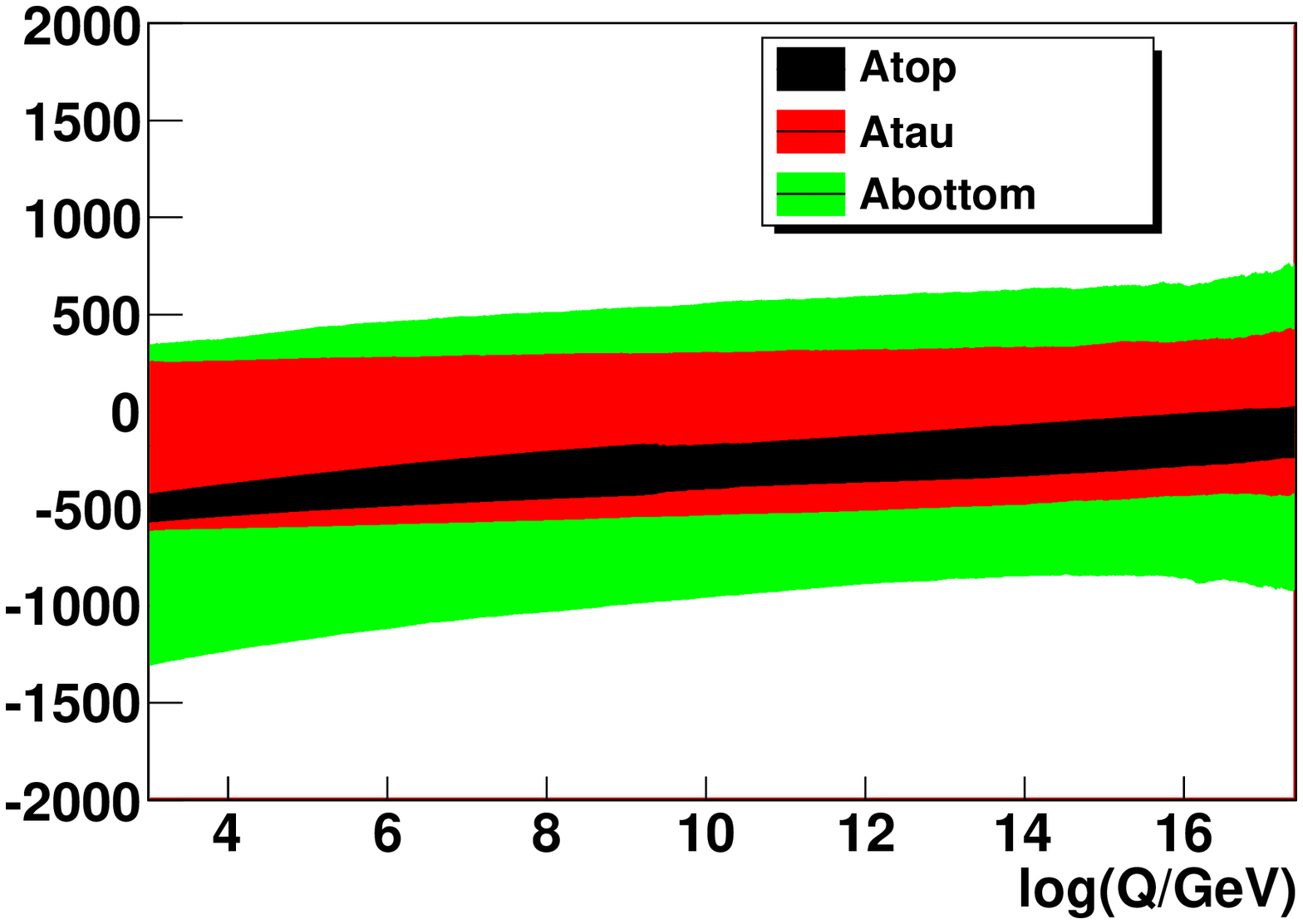}
  \includegraphics{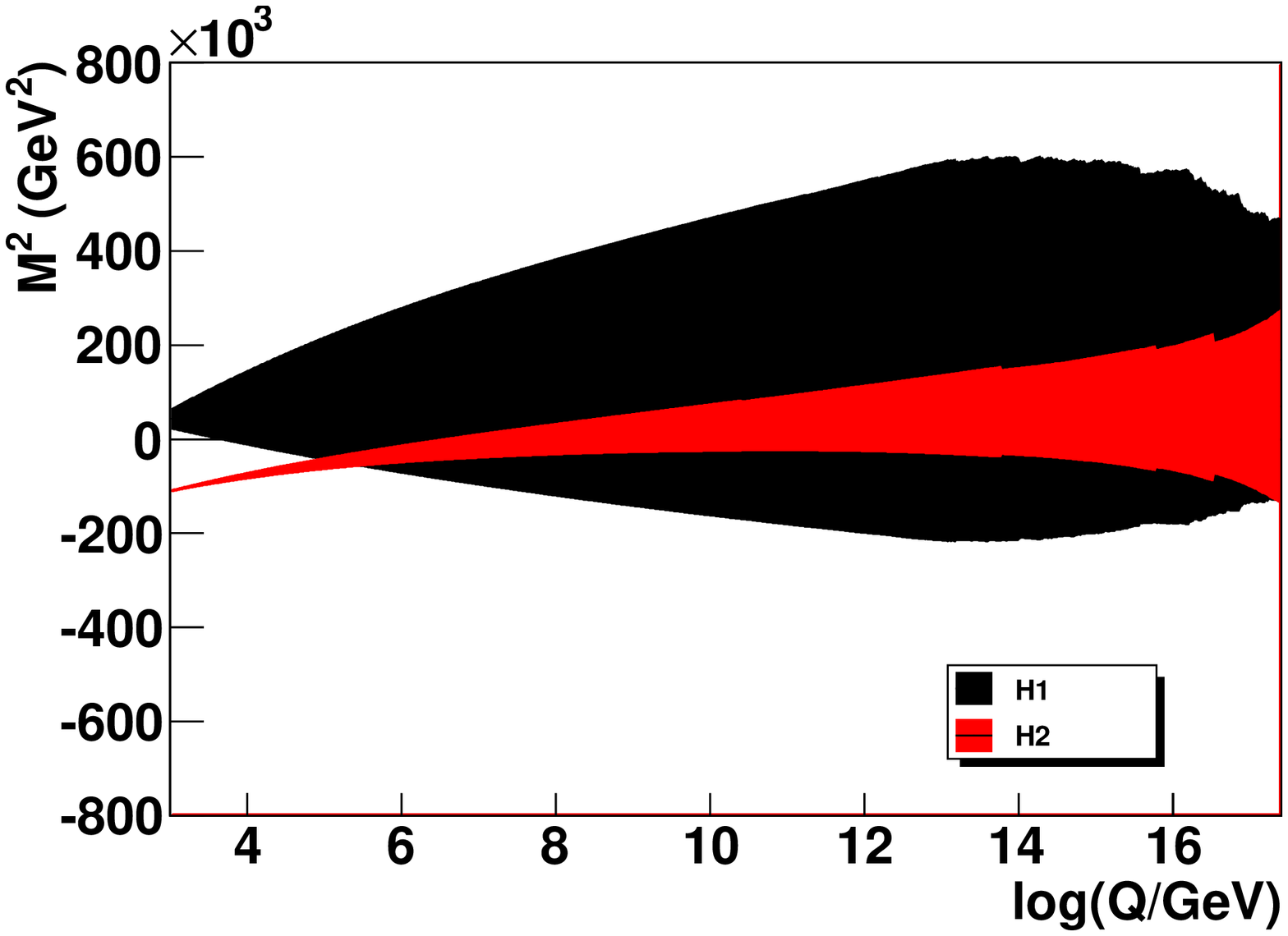}
}
\caption{(left) The evolution of the trilinear parameters of the third generation
is shown for the combined results from LHC+ILC. 
(right) The evolution of the squares of the Higgs sector parameters \MHOne\ and \MHTwo\ is shown
for LHC+ILC.}
\label{fig:LHCILCtrilinMH12}       
\end{figure}

Of the trilinear couplings, only \Atop\ can be measured with good precision with the 
mass measurements from the ILC and the edges from the LHC. Further measurements will be necessary
to constrain these parameters. For completeness sake the results are shown in the last line
of Table~\ref{tab:LHCILCunif} and in Figure~\ref{fig:LHCILCtrilinMH12} (left). 
Due to the large errors on the parameters determined at the electroweak scale, the parameters
are compatible with unification at all scales.
Fixing the unification scale to the one determined by the gaugino sector,
the error on \AZero\ is 200~GeV. This is a slight improvement 
compared to the error of 211~GeV on \Atop\ at the unification scale. Only
\Atau\ contributes to the reduction of the error as \Abottom\ has 
a larger error and is therefore irrelevant for the combination.

The evolution of the squares of the Higgs sector parameters \MHOne\ and \MHTwo, replacing the 
electroweak parameters $m_A$ and $\mu$ is shown
in Figure~\ref{fig:LHCILCtrilinMH12} (right). The square of \MHTwo\ is negative at the electroweak scale, 
as required by electroweak symmetry breaking, and reasonably well determined. As the evolution proceeds
to higher scales, the error increases significantly, somewhat faster for \MHOne. At the high scale the parameters
are compatible with the true value of $(100~\gev)^2$, but also with zero. Thus there is no significant contribution on the determination
of \mZero\ from these parameters.

The fixed scale results of the common parameters at the high scale 
can be compared to the mSUGRA parameter determination reported in Ref.~\cite{Lafaye:2007vs}.
While for the gauginos the error of the bottom-up determination compared 
to mSUGRA is about 2.5 times larger, the common scalar mass is determined with a 
precision 5 times better in mSUGRA. The largest difference is observed 
for the tri-linear coupling where the mSUGRA determination is
more precise than the bottom-up one by a factor of 20.
This shows that bottom-up and top-down do not lead to the 
same results.

The last question to be addressed is to determine the probability with 
which grand unification will be measured in the four measurements.
At the unification scale 95\% of the toy experiments show unification 
in the gaugino sector in agreement with the definition of the $\chi^2$ cut.
For the scalar parameters at the unification scale 
defined by the gauginos, the most precise measurement of this scale, 90\% unify. For the trilinear couplings 93\% of the 
toy experiments are compatible with grand unification.

The additional observables added by the ILC to the LHC dataset indeed increase the precision 
of the determination of the couplings in the gaugino sector by a large factor. Additionally 
different unification hypotheses can be tested (full scalar unification, separate unification 
for the light and heavy generations).

\subsubsection{Effect of threshold corrections at the high scale}

To study the effect of threshold corrections the measured value of 
\MThree\ is shifted by 3\% as in Section~\ref{sec:threshold}. 
The common gaugino mass is then determined to be 251.7~GeV and the scale is shifted by 0.05. 
The mass shift at the LHC alone is less than about half of a sigma. 
Due to the higher precision of the combination LHC+ILC, 
the shifts, while similar in absolute numbers, is now of the order of a sigma. 

The effect on the percentage 
of unifying parameter sets is also much larger: it decreases 
from 95\% to 77\%. For comparison, the effect is only half as large for the LHC alone.
The increased precision added by the ILC means that the effect of threshold effects become more important.

\subsubsection{Evolution with shifted data}

As in Section~\ref{sec:LHCsoftSUSY} to illustrate the effect of a shifted
dataset, the central values from SUSPECT are used, 
but the predictions as well as the evolution bottom-up are performed by SoftSUSY. 

In the gaugino sector the unification scale is determined 
with the same precision as before. The central values 
differ by less than 0.01, corresponding to one fifth of a sigma.  
\mOneHalf\  is determined to be 249.5~GeV, in excellent agreement 
with the determination using SUSPECT. The error on \mOneHalf\ is comparable at
1.6~GeV. 94\% of the parameter sets unify.

In the scalar sector 
\mZero\ is shifted by 1.6~GeV closer to the nominal value of SPS1a. The error
from the determination of 9~GeV is comparable to that from SUSPECT. 
Thus the shift is less than one fifth of a sigma. 

The study shows that the theoretical errors fulfill the requirement that they cover 
at least the difference between different calculations of the spectrum and the RGE running.
The results are in excellent agreement, showing that the analysis is robust.

\subsubsection{High scale MSSM}

\begin{figure}[htb]
\resizebox{0.5\textwidth}{!}{%
  \includegraphics{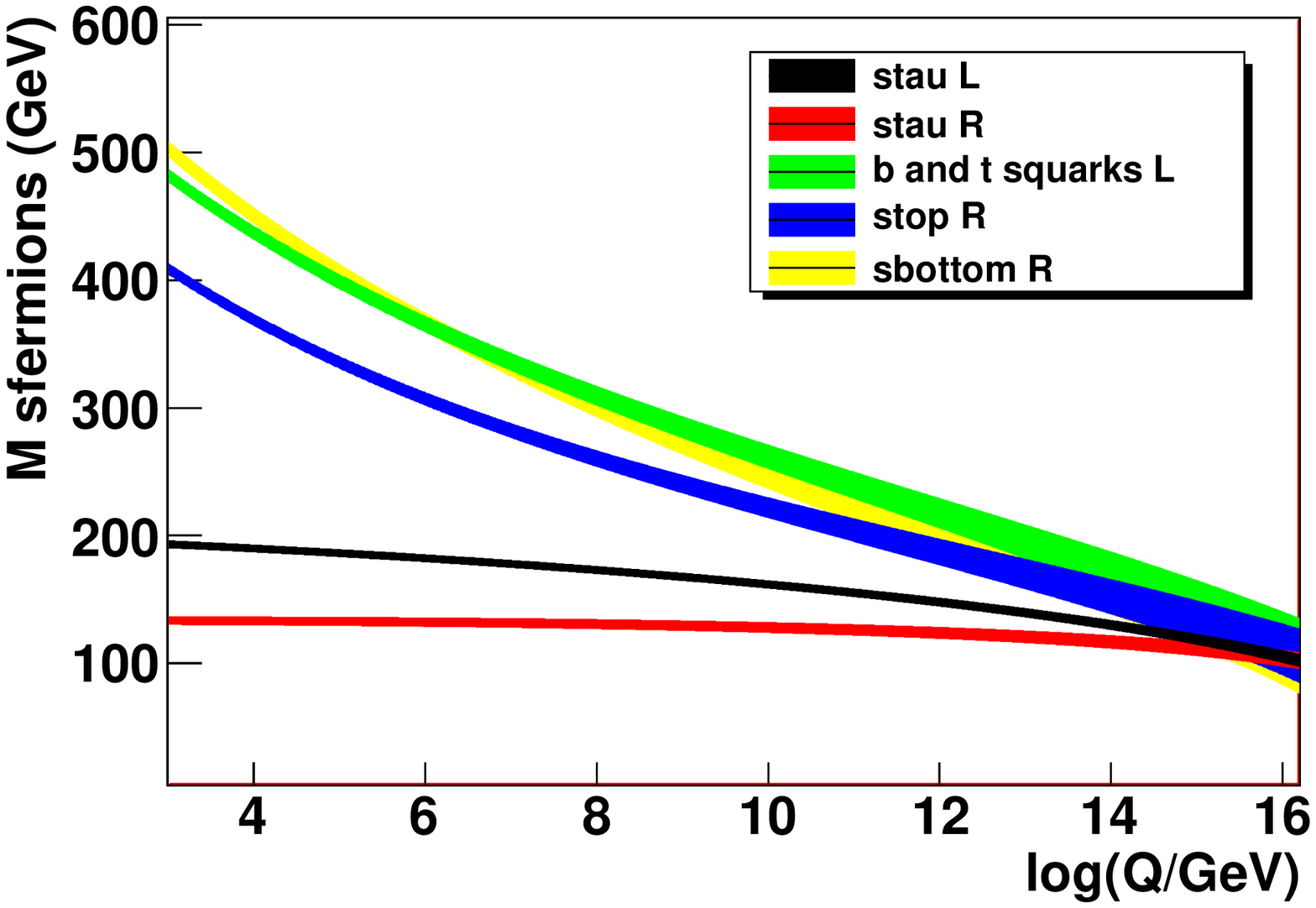}
  \includegraphics{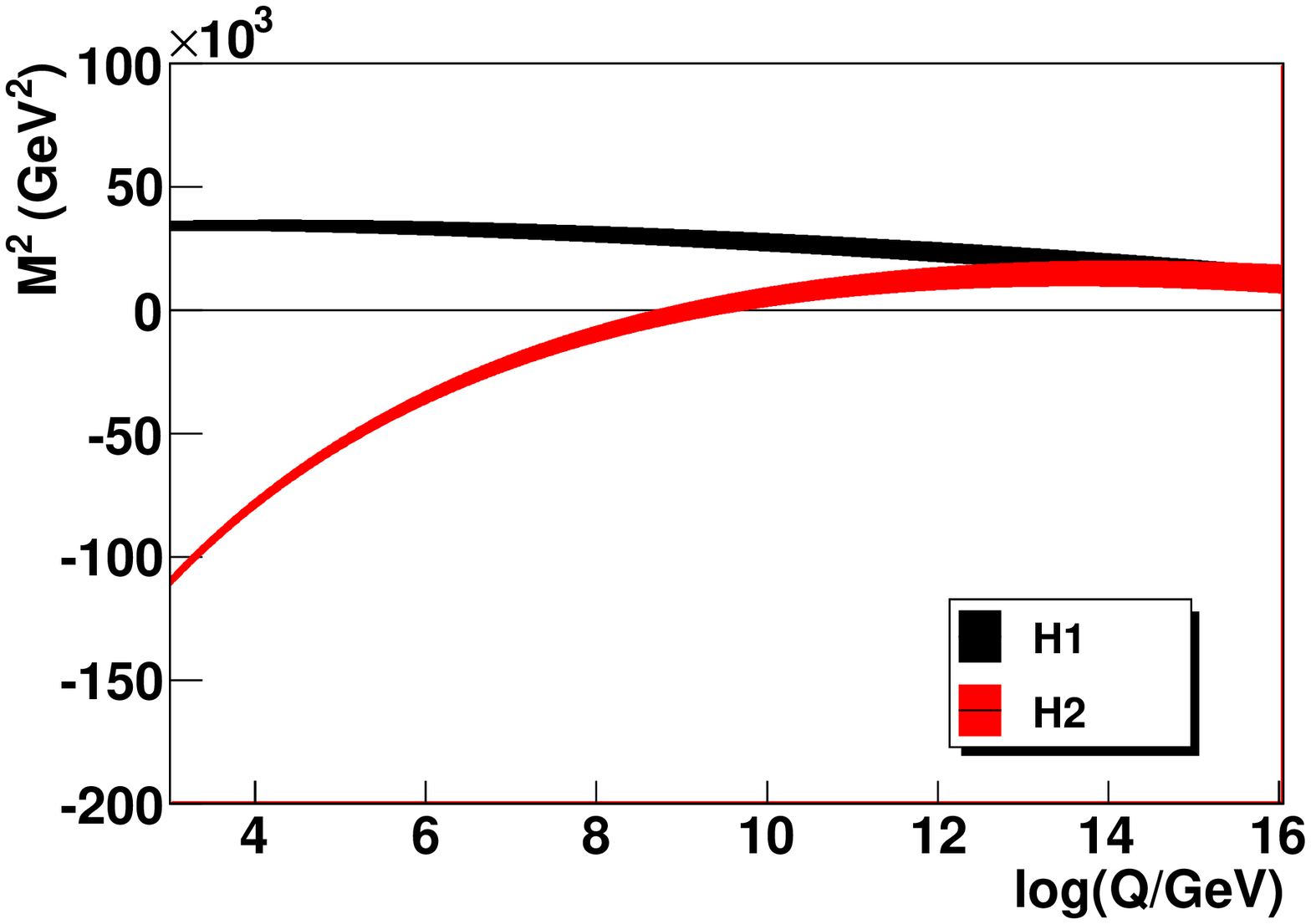}
}
\caption{The top-down evolution of high-scale MSSM parameters at the LHC+ILC is shown for : 
(left) third generation scalar mass parameters, (right) squares of the Higgs sector parameters \MHOne\ and \MHTwo.}
\label{fig:LHCILChigh}       
\end{figure}

As shown in Figure~\ref{fig:validEvents} (right), the fraction of
``non-tachyonic'' datasets is much higher in the LHC+ILC case.
But the comparison of the errors on the MSSM parameters evolved to the 
high scale with the high scale MSSM parameters shows again the difference
between a top-down and a bottom-up approach. The difference between the errors on 
the parameter determination at the unification scale is reduced due to the
increased precision added by the ILC measurements. In the gaugino sector 
these are now of the order of 10-20\%. In the scalar sector the differences remain
much larger, a factor 5 in the stau sector for example. 
The results of the top-down evolution shown in Figure~\ref{fig:LHCILChigh} can easily be compared with the bottom-up 
evolution shown in Figure~\ref{fig:LHCILCtrilinMH12} (right) and
Figure~\ref{fig:LHCILC3rdGen}.

The strong sensitivity to the Higgs mass term \MHTwo\, as illustrated in Figure~\ref{fig:LHCILChigh}, is not 
surprising since in most MSSM scenario its evolution, mainly driven by the top Yukawa coupling, 
drastically accelerates near the EW scale, where generally $\MHTwo^2$ changes sign. In other words 
the slope of its beta function becomes large around the EW breaking scale,
which is taken as the initial scale in a bottom-up evolution. Therefore in the bottom-up evolution even
a small error in the initial low scale \MHTwo\ value can induce a very large difference
at the high scale. In contrast, in a top-down evolution near the GUT scale
the initial slope of the beta function for \MHTwo\ is moderate, and the final value at the EW scale is 
less dependent on initial high scale boundary conditions, illustrating  a typical focusing
behavior.

To illustrate the impact of this behavior on the scalar mass determination, 
a bottom-up parameter set is selected for which \MselR, after 
evolution to the high scale, is about 60~GeV, i.e., far away from the SPS1a nominal values. 
The choice of \MselR\ to select the dataset/parameter set is motivated by the fact that 
it is well measured and \MselR\ is the lightest scalar mass. This parameter is therefore 
more likely to evolve to tachyon values in a bottom-up evolution.
A different choice of the scalar parameter departing from its SPS1a value would also have been possible.

Two high scale MSSM parameter determinations are performed with this dataset: one 
starting from the nominal SPS1a values and a second one 
starting from the MSSM results after evolution to the high scale. 
In the first case, the high scale MSSM scalar masses are determined at values within 2~GeV
of the SPS1a nominal values for the slepton masses. The
best-fit result of the second determination 
is compatible with the result of the bottom-up study, where apart from
\MselR\ other scalar masses are also far away from their nominal SPS1a values at the high scale.

\begin{figure}[ht]
\resizebox{0.5\textwidth}{!}{%
  \includegraphics{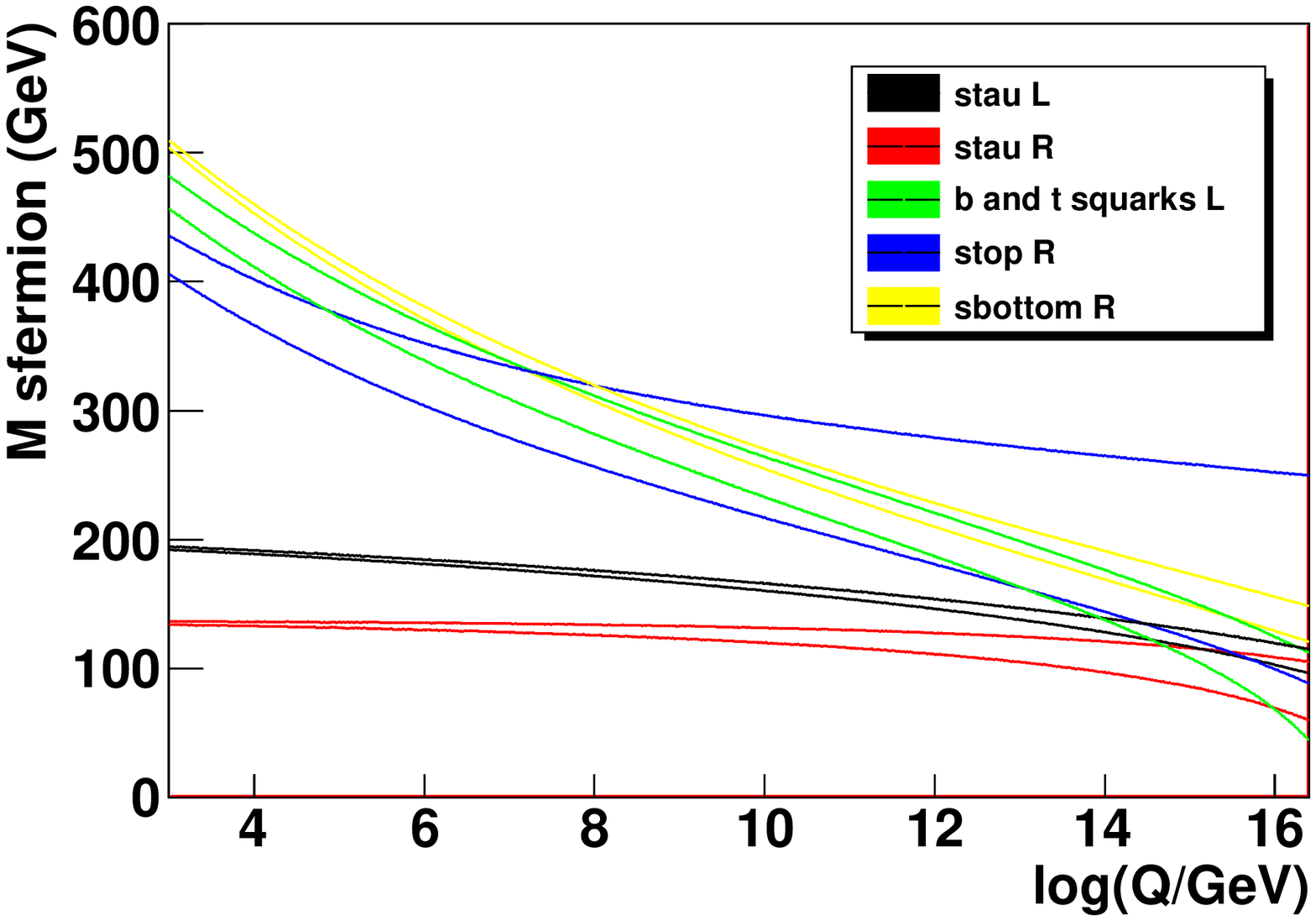}
  \includegraphics{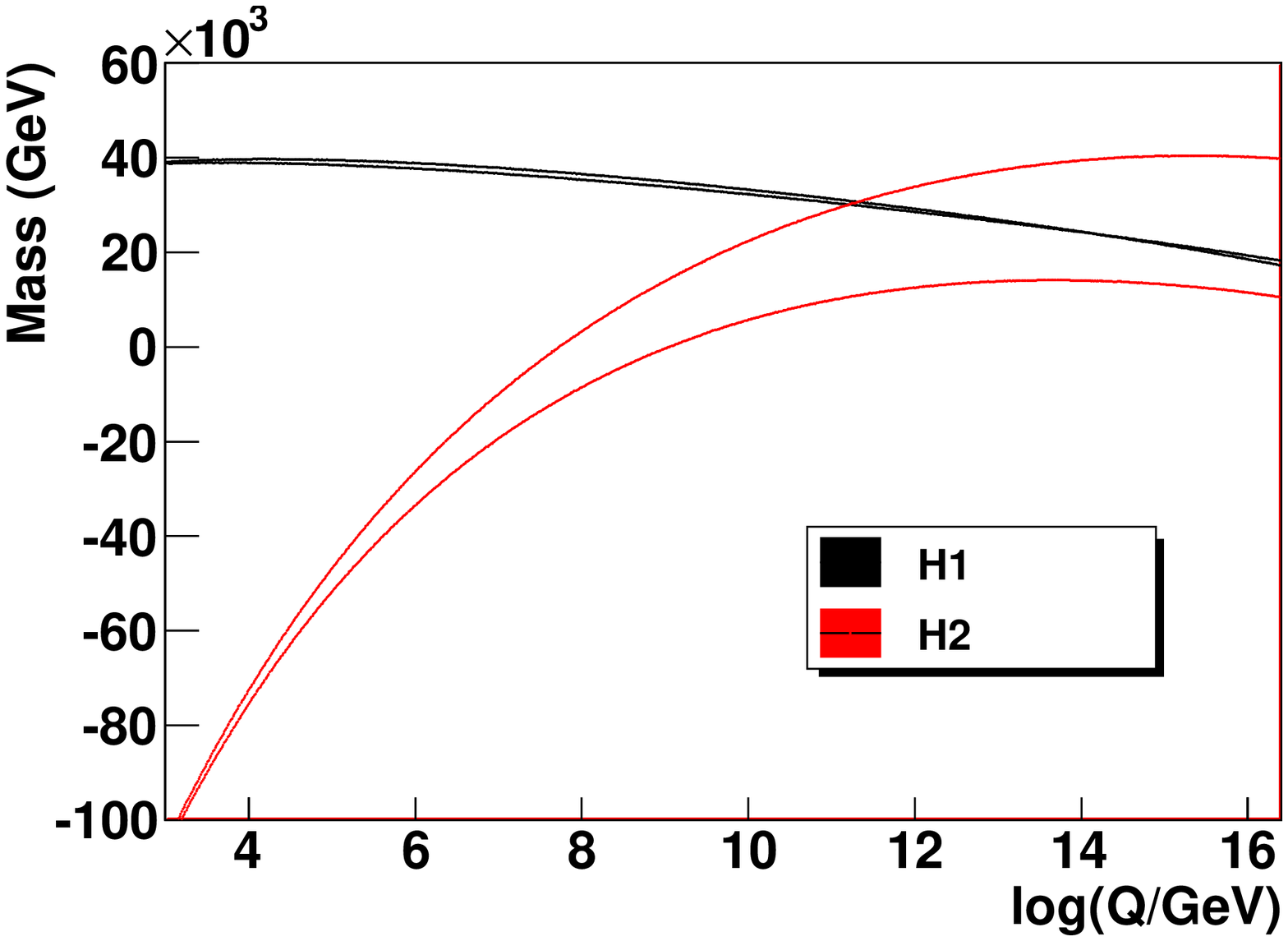}
}
\caption{The top-down evolution of high-scale MSSM parameters at the LHC+ILC is shown for : 
(left) third generation scalar mass parameters, (right) squares of the Higgs sector parameters \MHOne\ and \MHTwo.
The two lines for each parameter correspond to the two essentially degenerate solution for the same dataset. One
solution is close to the SPS1a values at the high scale, the other one far away. At the EW scale the mass parameters
with the exception of \MstR\ and \MsqThreeL\ converge to the same value.}
\label{fig:LHCILCfixedPoint}       
\end{figure}

The $\chi^2$ of both parameter determinations is very good, less than the degrees of freedom
with a slightly smaller value for the MSSM bottom-up result ($\Delta\chi^2=3/18$d.o.f.).
The result of the evolution of the two high scale MSSM determinations is shown in Figure~\ref{fig:LHCILCfixedPoint}
for the third generation scalar parameters and \MHTwo. All parameters with the exception
of \MstR\ and \MsqThreeL\ (and the tri-linear couplings) 
converge from extremely different values at the high scale to the same
value at the EW scale. This effect can be understood from the fact that 
at the leading one-loop RGE, both \MstR\ and \MsqThreeL\ evolution depend strongly 
on \MHTwo, which exhibits a strong variation around the EW scale as mentioned above. 
In contrast, other relevant scalar masses do not depend (at one-loop RGE level) 
on \MHTwo. Moreover the other scalar parameters RGE depend very little on \MstR\ and \MsqThreeL, the effect 
is suppressed 
by the bottom Yukawa coupling, which is quite small for SPS1a. 
Indeed, the evolution of the combination $2\cdot\MsqThreeL^2-\MstR^2$, which eliminates the dominant dependence on \MHTwo\ at
one--loop RGE, has essentially the same form for the two solutions as expected.

Given the smallness of $\Delta\chi^2$, the two solutions are de facto degenerate. It is interesting to
ask why the bottom-up solution has found an ever so slightly better solution than the standard high scale
study. While at the EW scale the difference of \MstR\ and \MsqThreeL\ for the two solutions is 
of the order of 20~GeV, i.e., close by, at the high scale the difference is of the order of 150~GeV. 
Thus at the EW scale the sampling of the parameter space is much easier: all mass parameters but 
two are the same and only a small excursion of 20~GeV in \MstR\ and \MsqThreeL\ is needed to 
find and differentiate between the two solutions. At the high scale however, all scalar parameters of the 
two solutions are far apart:
60GeV for \MselR, 150~GeV for \MstR\ etc. Therefore the parameter determination will find 
easily the solution close to the values of SPS1a, while it is more difficult to find the second solution
as a much larger parameter space has to be sampled. But even in this case, by construction, the top-down
extrapolation will miss the contribution of the tachyonic parameter sets from the bottom-up approach. 

The high scale MSSM top-down results presented in this study agree with the bottom-up results 
presented by other groups~\cite{Allanach:2004ud,Bechtle:2005vt}. In Refs.~\cite{Allanach:2004ud,Bechtle:2005vt}
additional ILC observables are used (polarized cross sections) and the theory
errors are set to zero. Additionally in Ref.~\cite{Allanach:2004ud} the 
trilinear parameters \Atau\ and \Abottom\ are required to be at the high scale 
compatible with \Atop\ within 2~sigma. This additional requirement leads 
mechanically to a reduction of the error on the corresponding mass parameters. 

\section{Conclusions}
The discovery of supersymmetry at the LHC will lead to a wealth of signatures 
which can be exploited to determine many MSSM parameters. In parameter regions similar to the SPS1a parameter point, they can be determined
at the LHC up to an at least 8-fold ambiguity in the gaugino sector. Although a part of those ambiguities
may be resolved, e.g., by a complementary study of the MSSM contributions to the dark matter relic density, 
a full resolution will likely require a complete 
observation of the sparticle spectrum at the ILC. 

Starting from the electroweak scale, we can test the unification of different supersymmetry-breaking parameters. 
While remaining ambiguities make it impossible to measure unification at the LHC, 
it will nonetheless be possible to classify
solutions into the ones compatible and the ones not compatible
with unification. In the case of an ambiguous solution (DS1) which differs from the
true solution only by the sign of $\mu$, the differentiation will be difficult
as about 38\% of the parameter sets corresponding to this wrong solution nevertheless unify.

This way, at the LHC the unified gaugino mass parameter can be measured bottom-up to 
about 2\% and the logarithm of the unification scale to 1.7\%. 
Adding the ILC data improves the determination of the mass by more than a factor~3 
and the unification scale by almost one order of magnitude. 
In the scalar sector the errors are generically larger at the level of 
10\%. The errors on the trilinear couplings are too large to be used
for a determination of the unification scale.

The robustness of our results we have confirmed by comparing two 
different renormalization group tools: SUSPECT and SoftSUSY. The parameter determination as well the 
evolution are in good agreement within the errors. 
Threshold corrections at the high scale were studied for a particular model, motivated by 
SU(5) grand unification. The percentage of parameter sets unifying is affected
more strongly including the ILC observables providing an increased sensitivity. 

Finally, our study show that a proper bottom-up approach will clearly lead to different 
results from simply determining the parameters of the high scale MSSM (or mSUGRA). 
In addition to resolving the ambiguities at the LHC, the ILC plays a strong role in the stabilization 
of the validity of the parameter sets as
function of the scale.

\section*{Acknowledgments}

Part of this work was developed in the GDR Terascale of the CNRS.
M.R. acknowledges support by the Deutsche Forschungsgemeinschaft via the
Sonderforschungsbereich/Transregio SFB/TR-9 ``Computational Particle
Physics'' and the Initiative and Networking Fund of the Helmholtz
Association, contract HA-101 (``Physics at the Terascale''). 
We would like to thank P.~Zerwas and W.~Porod for clarifying the
definition of their observables 
and the assumptions used in their analysis.

%
%

\end{document}